\documentclass[usenatbib,onecolumn]{mn2e}

\usepackage{graphicx}
\usepackage{color}
\usepackage{multirow}
\usepackage{amssymb}
\usepackage{amsmath}
\usepackage{anysize}
\usepackage{aas_macros}
\usepackage[totalwidth=480pt, totalheight=680pt]{geometry}

\renewcommand{\vec}[1]{\bmath{#1}}

\newcommand{\DS}{\displaystyle}

\newcommand{\mathi}{\rm i}

\newcommand{\av}[1]{\left<#1\right>}

\title[Instabilities in Relativistic Jets]
      {Linear stability analysis of magnetized  relativistic jets: the nonrotating case}

\author[G.Bodo et al.]
       {G. Bodo$^{1}$\thanks{E-mail:
bodo@oato.inaf.it}, G. Mamatsashvili$^{2}$, P. Rossi$^{1}$ and A. Mignone$^{3}$\\
$^{1}$INAF/Osservatorio Astronomico di Torino, Strada Osservatorio 20, 10025 Pino Torinese, Italy\\
$^{2}$ Department of Physics, Faculty of Exact and Natural Sciences, Tbilisi State University, Il. Chavchavadze ave. 3, Tbilisi 0179, Georgia\\
$^{3}$Dipartimento di Fisica Generale ``Amedeo Avogadro'' Universit\`a degli Studi di Torino, Via Pietro Giuria 1, 10125 Torino, Italy}

\begin{document}

\date{Accepted ??. Received ??; in original form ??}

\pagerange{\pageref{firstpage}--\pageref{lastpage}} \pubyear{2007}

\maketitle

\label{firstpage}

\begin{abstract}
We perform a linear analysis of the stability of a magnetized relativistic non-rotating cylindrical flow in the aproximation of zero thermal pressure, considering only the $|m| = 1$ mode.   We find that there are two modes of instability: Kelvin-Helmholtz and current driven. The Kelvin-Helmholtz mode is found at low magnetizations and its growth rate  depends very weakly on the pitch parameter. The current driven modes are found at high magnetizations and the value of the growth rate and the wavenumber of the maximum increase as we decrease the pitch parameter. In the relativistic regime the current driven mode is splitted in two branches,  the branch at high wavenumbers is characterized by  the eigenfunction concentrated in the jet core,  the branch at low wavenumbers is instead characterized by the eigenfunction that extends outside the jet velocity shear region. 
\end{abstract}


\section{Introduction}
%
%
%

The formation and propagation of astrophysical jets are strongly affected by plasma instabilities, whose study is therefore  
of fundamental importance for  understanding their dynamics and their associated phenomenology. 
In jets there are several possible different kinds of instability, among them the most studied are the  Kelvin-Helmholtz instability (KHI) driven by the velocity shear  between the jet and the ambient medium  and the current driven instability (CDI) associated with a longitudinal current and therefore with the toroidal component of magnetic field. 
Since the most promising models for the acceleration and collimation of jets involve the presence of a magnetic field with footpoints anchored to a rotating object (an accretion disk or a spinning star or black hole), the presence of a toroidal field component is a natural consequence and CDI may play an important role in the jet propagation.  Among CDI, the $|m| =1$ kink mode is the most effective, leading to an helical displacement  of the whole jet body, and for this reason is the only one we will consider in this paper. Recent  numerical simulations  have shown the development of such instabilities both in Newtonian  \citep{Moll08, Nakamura04} and in relativistic  \citep{McKinney09, Mignone10} jets.  An important step towards a better understanding of simulation results is a linear analysis of the instabilities, that is still largely missing for the relativistic magnetohydrodynamic (MHD) regime. 

KHI  have been extensively studied  in several different configurations  in the Newtonian \citep[see e.g.][]{Bodo89, Birkinshaw91, Hardee92, Bodo96, Hardee06} and relativistic \citep[see e.g.][]{Ferrari78, Hardee79,  Urpin02, Perucho04, Perucho10} cases, but very few linear analyses have been presented for a relativistic magnetized jet \citep{Hardee07}.     Similarly,  CDI  have been widely studied in the Newtonian limit \citep[see e.g.][]{Appl92, Appl96, Begelman98, Appl00, Baty02}, however, for the relativistic MHD case, only the force-free limit has been considered \citep{Pariev94, Pariev96, Lyubarski99, Tomimatsu01, Narayan09}. \citet{Pariev94, Pariev96} have considered the case in which the longitudinal magnetic field is constant, showing that in this case the jet is stable, on the contrary  \citet{Lyubarski99}  showed that jets with longitudinal magnetic field decreasing outward can be unstable.  \citet{Tomimatsu01}  finally derived a general necessary condition for instability.  

The force-free limit is valid when the energy density of electromagnetic fields is much larger than the energy density of matter, i.e. when the jet energy flux is mainly in the form of Poynting flux.  Acceleration models predict that jets start being Poynting dominated and progressively   undergo   a transition to a matter dominated state, but how fast the transition occurs is uncertain. From the observational point of view, the evidence is that, above  the parsec scale,  jets cannot be Poynting dominated \citep{Sikora05, Celotti08} .  On the other hand, it has been suggested \citep{Sikora05} that jet instabilities may play a role in this conversion and that blazar activity could be linked to the development of such instabilities either through shock formation or through direct magnetic energy dissipation processes. These considerations suggest that it is very important to go beyond the force-free limit  in the instability analysis and  this is exactly the aim of the present paper in which we will analyze the linear stability properties of a cylindrical magnetized relativistic flow, taking into account  the effects of matter inertia, but still neglecting thermal pressure.  Since the equilibrium configurations of the jet may be quite complex, with many possible sources of instabilities, like the longitudinal jet velocity, the toroidal field and the jet rotation, we will not start from a full configuration  where all the above elements are present, but  from a simpler case where one of the above elements, namely rotation, is absent.  The additional effects introduced by rotation will then be examined in a following paper.   In section 2 we will describe the physical problem, the relevant equations, the general equilibrium configuration and the relevant parameters, while in section 3 we will derive the linearized equations and describe the procedure for finding the normal modes.  In section 4 we will present the results of our analysis for the non rotating case, first for a static configuration and then for a moving jet. Finally in section 5 we will summarize our findings.

\section{Problem Description}
%
%
%

We study the stability of a cold (zero pressure) relativistic magnetized  cylindrical flow. The relevant equations are continuity and momentum coupled with Maxwell  equations:
\begin{equation}\label{eq:drho/dt}
\frac{\partial   }{\partial t} (\gamma \rho) + \nabla \cdot (\gamma \rho  \vec{v}) = 0 \,,
\end{equation}
\begin{equation}\label{eq:dm/dt}
\gamma \rho \frac{\partial}{\partial t} (\gamma \vec{v} ) + \gamma \rho (\vec{v} \cdot \nabla ) (\gamma \vec{v} ) = \vec{J} \times \vec{B} + \frac{( \nabla \cdot \vec{E} ) \vec{E}}{4 \pi}  \,,
\end{equation}
\begin{equation}\label{eq:dB/dt}
\frac{\partial   \vec{B}}{\partial t} = - \nabla \times \vec{E}   \,,
\end{equation}
\begin{equation}\label{eq:dE/dt}
\frac{\partial   \vec{E}}{\partial t} =  \nabla \times \vec{B} - 4 \pi \vec{J} 
  \,,
\end{equation}
where $\rho$ is the proper density, $\gamma$ is the Lorentz factor,  and $\vec {v}$, $\vec {B}$, $\vec{E}$, $\vec{J}$ are respectively the velocity, magnetic field, electric field and current 3-vectors. The units are chosen so that the speed of light is $c = 1$, we also remark that in the following a factor of $\sqrt{4 \pi}$ will be reabsorbed in the definitions of $\vec{E}$ and $\vec{B}$. The first step in the stability analysis is to define an equilibrium state satisfying the stationary form of Eqs. (\ref{eq:drho/dt}- \ref{eq:dE/dt}) and this will be done in the next subsection.

\subsection{Equilibrium Configuration}
%
%
\label{sec:equilibrium}

We adopt a cylindrical system of coordinates $(r,\varphi,z)$ (with versors $\vec{e_r}, \; \vec{e_\varphi}, \; \vec{e_z}$)
and seek for  axisymmetric steady-state solutions for a relativistic magnetized jet,
i.e., $\partial_t = \partial_\varphi = \partial_z=0$.
We assume that the jet propagates in the vertical ($z$) direction and
the magnetic field configuration consists of a vertical (poloidal) component 
$B_z$ and a toroidal component $B_\varphi$ and can be expressed as
\begin{equation}
  \vec{B} = B_\varphi(r) \vec{e_\varphi} +  B_z(r) \vec{e_z} \,.
\end{equation}
The magnetic field configuration can be characterized by the pitch parameter
\begin{equation}\label{eq:pitch}
  P = \frac{r B_z}{B_\varphi} \,.
\end{equation}
From the stationarity condition, the continuity equation and the independence of $v_r$ on $z$ and $\phi$, we get $v_r = 0$ and 
from the condition $\nabla\times\vec{E}=0$  we obtain that the velocity can then be written as
\begin{equation}
\vec{v} = v_z(r) \vec{e_z} + v_\varphi(r) \vec{e_\varphi} = \kappa(r) \vec{B} + \Omega(r) r \vec{e_\varphi}  \,,
\end{equation}
where $v_\varphi$ is the fluid toroidal velocity and $\Omega$ is the angular velocity of field lines and they are related by
\begin{equation}
 \Omega = \frac{v_\varphi}{r} - \frac{v_z B_\varphi}{r B_z} 
        = \frac{v_\varphi}{r} - \frac{v_z}{P} \,.
\label{eq:Omega}
\end{equation}
The electric field is always directed radially and can be expressed as 
\begin{equation}
\vec{E} = - \Omega r B_z  \vec{e_r}\,.
\end{equation}

The only remaining non-trivial equation is given by the radial component of the momentum equation (\ref{eq:dm/dt}) which simplifies to
\begin{equation}\label{eq:radial_eq}
  \rho\gamma^2v_\varphi^2 = \frac{1}{2r}\frac{d(r^2H^2)}{dr} 
  + \frac{r}{2}\frac{dB_z^2}{dr} \,,
\end{equation}
where $H^2 = B_\varphi^2 - E_r^2$.  
In the nonrelativistic limit $H$ reduces to $B_\varphi$ and the equilibrium condition acquire the classical Newtonian form
\begin{equation}\label{eq:radial_eqMHD}
  \rho v_\varphi^2 = \frac{1}{2r}\frac{d(r^2B_\varphi^2)}{dr} 
  + \frac{r}{2}\frac{dB_z^2}{dr} \,.
\end{equation}
Eqs. (\ref{eq:radial_eq}) or (\ref{eq:radial_eqMHD}) leave the freedom of choosing the radial profiles of all flow variables but one and then solve for the remaining profile.  
We begin by prescribing the profiles of the proper density and Lorentz factor that well describe a jet configuration, with the velocity and density variations concentrated inside the jet radius $r_j$
\begin{eqnarray}\label{eq:rho_prof}
 \rho(r) &=&\DS \eta + \frac{1-\eta}{\cosh(r/r_j)^6} \,,
\\ \noalign{\medskip}
\label{eq:vz_prof}
 \gamma_z(r) & = & 1 + \frac{\gamma_c - 1}{\cosh(r/r_j)^6} \,,  
\end{eqnarray}
where $\eta$ is the ambient/jet density contrast, $\gamma_z(r)$ is the Lorentz factor relative to the $z$ component of the velocity only, while $\gamma_c = 1/\sqrt{1-v_c^2}$ is the Lorentz factor on the axis where the vertical flow velocity is $v_z(0) = v_c$.
From now on, we will use the subscript $c$ to denote values at $r=0$.

We note that, in the Newtonian case, the presence of a longitudinal velocity has no effect on the radial equilibrium Eq. (\ref{eq:radial_eqMHD}), while it changes the relativistic Eq. (\ref{eq:radial_eq}) modifying the centrifugal term. 

In the Newtonian limit, it is then customary to prescribe the profile of the azimuthal field $B_\varphi$ and this choice is more arbitrary since we have no direct information about the magnetic configuration in astrophysical jets. 
The choice of the $B_\varphi$ distribution is equivalent to a choice of the distribution of the longitudinal component of the current and also determines the behavior of the pitch parameter $P(r) $, that is important for the stability properties. 
In principle, one can then have several equilibria characterized by different forms of the current distribution, that can be  more or less concentrated, can peak on the axis or at the jet boundary,  and  can close in different ways \citep[see e.g.][]{Appl00, Bonanno08, Bonanno11}.  
In this limit, we start by considering the azimuthal field profile
\begin{equation} \label{eq:Bphi_prof}
 B_\varphi^2 = \frac{B_{\varphi c}^2}{(r/r_j)^2}\left[1 - \exp\left(-\frac{r^4}{a^4}\right)\right]\,,
\end{equation}
where $a$ is the magnetization radius and $B_{\varphi c}$ determines the maximum field strength. This profile  corresponds to a current distribution peaked on the jet axis and, choosing $a < r_j$,  concentrated inside the jet. 
In addition, one can assume that the current closes at very large distances from the jet.
In the relativistic limit the most natural generalization is to prescribe in a similar way the behavior of $H$
\begin{equation} \label{eq:H2_prof}
 H^2 = \frac{H^2_c}{(r/r_j)^2}\left[1 - \exp\left(-\frac{r^4}{a^4}\right)\right]\,.
\end{equation}
We observe that, in the absence of rotation, $H$ represents the azimuthal field strength measured in the proper frame. Prescribing the profile of $H$ instead of $B_\varphi$ modifies the current profile in the laboratory frame, introducing a return current in the region of the velocity shear  and this has consequences for the stability properties as it will be discussed below.

Furthermore, the equilibrium configuration may be modified by the presence of rotations to different degrees: in one extreme case the gradient of $r^2 H^2$ in Eq. (\ref{eq:radial_eq}) (or the gradient of  $r^2 B_\varphi^2$ in Eq. (\ref{eq:radial_eqMHD}))  is exactly balanced by the centrifugal force and $B_z$ is constant, in the other extreme it is balanced by the gradient of $B_z^2$.  
This suggests to introduce, more generally, a parameter $\alpha\in[0,1]$ so that the equilibrium poloidal magnetic field is given by 
\begin{equation}\label{eq:Bz_prof}
  B^2_z = B^2_{zc} - (1 - \alpha) \frac{H^2_c\sqrt{\pi}}{(a/r_j)^2}{\rm erf}
   \left(\frac{r^2}{a^2}\right)
\end{equation}
where $\mathrm{erf}$ is the error function, $\alpha = 0$ corresponds to the absence of rotation, while $\alpha = 1$ corresponds to maximum rotation. 
Introducing the expression for $B_z$ given by Eq. (\ref{eq:Bz_prof}) in Eq. (\ref{eq:radial_eq})   we can get the azimuthal velocity from  
\begin{equation}\label{eq:vphi_prof}
  2\rho\gamma^2v_\varphi^2 = \frac{\alpha}{r}\frac{d(r^2H^2)}{dr}
\end{equation}
Finally the azimuthal field is obtained from the definition of $H^2$ using $E_r=-(v_\varphi B_z - v_zB_\varphi)$. 
This yields a quadratic equation in $B_\varphi$ with solution
\begin{equation}\label{eq:Bphi}
B_\varphi = \frac{-v_\varphi v_zB_z\mp\sqrt{v_\varphi^2B_z^2 + H^2(1-v_z^2)}}{1-v_z^2} \,.
\end{equation}
Here we consider the negative branch because it guarantees that $B_\varphi$ and $v_\varphi$ have opposite signs, as suggested by acceleration models.
Thus, in our model the radial profile of the pitch parameter, Eq. (\ref{eq:pitch}), is always negative.
We choose to control the magnetic field configuration by specifying the value of the pitch  on the axis $P_c$ and the ratio between the energy density of the matter and the magnetic energy density $M^2_A$, where
\begin{equation}\label{eq:P_cnd_MA}
  P_c \equiv \left|\frac{rB_z}{B_\varphi}\right|_{r=0} \,,\qquad
  M_a^2 \equiv \frac{(\rho \gamma_c^2)}{\av{\vec{B}^2}} \,, 
\end{equation}
and $\av{\vec{B}^2}$ represents the average across the beam:
\begin{equation}\label{eq:Bav}
  \av{\vec{B}^2} = \frac{\int_0^{r_j} (B_z^2 + B_\varphi^2)r\,dr}
                        {\int_0^{r_j} r \,dr}\,.
\end{equation}
We note that $P_c>0$ by construction although the radial profile of the pitch parameter (Eq. \ref{eq:pitch}) using the negative branch of Eq (\ref{eq:Bphi}) in the definition of $B_\varphi$ is always negative.
The constants $B_{zc}^2$ and $H_c^2$ appearing in the above equations can be found in terms of $P_c$ and $M_a$ from the simultaneous solution of the two expressions given in Eq. (\ref{eq:P_cnd_MA}), together with (\ref{eq:Bav}). 
In particular, from the definition of the pitch parameter, after some algebra we find (in the $r\to0$ limit)
\begin{equation}\label{eq:Bz0}
  a^4B_{zc}^2 = \frac{H_c^2P_c^2}{1 - (H_cP_c\sqrt{\zeta} - v_z)^2} 
\end{equation}
where $\zeta = 2 \alpha/(\gamma_c^2\rho a^4)$.
In practice, Eq (\ref{eq:Bz0}) is used to compute $B_{zc}^2$ for trial values
of $H_c^2$ until the prescribed value of magnetic energy  is satisfied.

Summarizing, our equilibrium configuration depends on the 5 parameters $\eta$, $\gamma_c$, $\alpha$, $P_c$ and $M^2_a$ specifying, respectively, the jet density contrast, bulk flow velocity, strength of centrifugal force, magnetic pitch and the ratio between the energy density of the matter and the  magnetic energy density.

\section{Linearized Equations}
\label{sec:linequations}

Let us consider small perturbations  $\rho_1, \vec{v}_1,
\vec{B}_1, \vec{E}_1$ to the equilibrium state described
above, which hereafter will be identified by a zero subscript. The linearized continuity, momentum, induction equations and the ideal MHD conditions are
\begin{equation}\label{eq:lincont}
\frac{\partial}{\partial t}(\gamma_0\rho_1+\gamma_1\rho_0)
 +\nabla\cdot(\gamma_0\vec{v}_0\rho_1+\gamma_1\vec{v}_0\rho_0
               +\gamma_0\rho_0\vec{v}_1)=0 \,,
\end{equation} 
\begin{eqnarray}\label{eq:linmom}
\rho_0\gamma_0\left(\frac{\partial}{\partial t}
  +\vec{v}_0\cdot\nabla\right)(\gamma_1\vec{v}_0
      +\gamma_0\vec{v}_1)+ \rho_0(\gamma_1\vec{v}_0+\gamma_0\vec{v}_1)\cdot\nabla(\gamma_0\vec{v}_0)+
\rho_1\gamma_0\vec{v}_0\cdot\nabla(\gamma_0\vec{v}_0)=\nonumber
\\=
(\nabla\times\vec{B}_0)\times\vec{B}_1+(\nabla\times\vec{B}_1)\times\vec{B}_0
    +\vec{B}_0\times\frac{\partial \vec{E}_1}{\partial t}
+\vec{E}_1(\nabla\cdot\vec{E}_0)+\vec{E}_0(\nabla\cdot\vec{E}_1)\,,
\end{eqnarray}
\begin{equation}\label{eq:linind}
  \frac{\partial \vec{B}_1}{\partial t}=-\nabla\times\vec{E}_1 \,,
\end{equation}
\begin{equation}\label{eq:linMHD}
\vec{E}_1=-\vec{v}_1\times\vec{B}_0-\vec{v}_0\times\vec{B}_1\,.
\end{equation}

Assuming now the perturbations to be of the form $\propto \exp \left( {{\mathi}} \omega t - {{\mathi}} m \varphi - {{\mathi}} k z \right)$,  after  lengthy algebraic manipulations described in the Appendix \ref{ap:linear}, we arrive at a system of two first order differential equations in the radial coordinate for the two basic variables -- the radial displacement  $\xi_{1r}$ and the perturbed electromagnetic pressure $\Pi_1$  defined as
\begin{equation}
  \xi_{1r}=-{\mathi} v_{1r}/\tilde{\omega}\,,
\end{equation}
and 
\begin{equation}
\Pi_1=\vec{B}_0\cdot\vec{B}_1-\vec{E}_0\cdot\vec{E}_1=B_{0\varphi}B_{1\varphi}+B_{0z}B_{1z}-E_{0r}E_{1r} \,.
\end{equation}
The system of equations can then be written as
\begin{equation} \label{eq:dxi/dr}
\left. D\frac{d\xi_{1r}}{dr}=\left(C_1+\frac{C_2-D
k_B'}{k_B}-\frac{D}{r}\right)\xi_{1r} - C_3 \Pi_1 \right.
\end{equation}

\begin{multline}
D\frac{d\Pi_1}{dr} = \left[A_1
D-\frac{\rho_0\gamma_0^2v_{0\varphi}^2}{r}\left(C_1+\frac{C_2-D
k'_B}{k_B} \right) + \frac{C_4}{r}  + C_5 \right]\xi_{1r}+
\\+
\frac{1}{r}\left(\rho_0\gamma_0^2v_{0\varphi}^2C_3-2D + \frac{C_6}{r} \right)+ C_7 \Pi_1
\label{eq:dp/dr}
\end{multline}
where 
\begin{equation}
D=(\sigma+1)B_0^2\tilde{\omega}^2+\sigma k_B\left[2\tilde{\omega}(\vec
{v}_0\cdot\vec{B}_0)-\frac{k_B}{\gamma_0^2}\right]\,,
\end{equation}
\begin{equation}
\sigma =\frac{B_0^2}{\rho_0\gamma_0^2}, \qquad \tilde{\omega} \equiv  \omega-\frac{m}{r}v_{0\varphi}-kv_{0z} \,.
\end{equation}
The quantities $A_1, A_2, C_1, C_2, C_3, C_4, C_5, C_6, C_7, C_8$ depend on the chosen profiles of the equilibrium solution and are given in the Appendix  \ref{ap:coefficients}.
These two equations together with the appropriate boundary conditions represent an eigenvalue problem, where $\omega$ is the eigenvalue (we observe that we here adopt a temporal approach to the stability analysis). 
We have instability when $\omega$ has a negative imaginary part.

The domain of integration for Eqs.  (\ref{eq:dxi/dr}) and (\ref{eq:dp/dr}) cover the interval from $0$ to $\infty$, so we have to specify the boundary conditions at $r=0$ and for $r \rightarrow \infty$. 
On the axis at $r=0$ the equations are singular but the solutions have to be regular while at infinity the solutions have to decay and no incoming wave is allowed (Sommerfeld condition).
For finding the eigenvalue we use a shooting method with a complex secant root finder. 
The numerical integration cannot start at $r=0$ (because of the singularity), so we start at a small distance from the origin where the solution is obtained through a series expansion of the equations described in the  Appendix \ref{ap:small_r}. 
Similarly, we start a backward integration from a sufficiently large radius, where the asymptotic solution is obtained as described in the Appendix \ref{ap:large_r} and then we match the two numerical solutions at an intermediate radius. 
Furthermore, we have to consider that Eqs.  (\ref{eq:dxi/dr}) and (\ref{eq:dp/dr})  may have singular points, that arise  in the following  cases:
\begin{enumerate}
  \item when the Doppler shifted frequency, $\tilde{\omega}$, appearing in the denominators in $A_1$ and $A_2$,  becomes zero at some $r=r_c$, $\tilde{\omega}(r_c)=0$. 
  This is a well-known corotation singularity, when the phase speed of a wave perturbation coincides with the basic flow velocity at $r=r_c$ and corresponds to
resonant interaction between waves and the background jet flow. 
This corotation singularity is a physical one.

  \item when the determinant becomes zero at some $r=r_A$, $D(r_A)=0$. 
  The latter condition is a quadratic equation with respect to $\tilde{\omega}$ that gives
\begin{equation}
\widetilde{\omega} = \frac{k_B}{\gamma^2_0} \cdot \frac{1}{\left.
\left( \vec{v}_0 \cdot \vec{B}_0 \right) \pm \sqrt{\left(
\vec{v}_0 \cdot \vec{B}_0 \right)^2 + \frac{B^2_0
+ \rho_0 \gamma^2_0}{\gamma^2_0}} \right.} \,,
\end{equation}
from which we derive a  phase speed equal to that of a relativistic Alfv\'en wave (see Keppens \& Meliani 2008, eq. 28, Istomin \& Pariev 1996) 
\begin{equation}
v_{\rm ph} = \hat{\vec{n}} \cdot \vec{v}_0 +
   \frac{\hat{\vec{n}} \cdot \vec{B}_0}{\gamma^2_0}  \frac{1}{\left(
   \vec{v}_0 \cdot \vec{B}_0 \right) \pm \sqrt{\rho_0 + 2
   p_{\rm mag}}} \,,
\end{equation}
where $\hat{\vec{n}}=(k^2+m^2/r^2)^{-1}(m/r, k)$ is the versor of the wave vector and
\begin{equation}
   p_{\rm mag} = \frac{B^2_0}{\gamma^2_0} + \left(
   \vec{v}_0 \cdot \vec{B}_0 \right)^2 
\end{equation}
is the total pressure.
In other words, the above determinant is zero and therefore equations (\ref{eq:dxi/dr}) and (\ref{eq:dp/dr}) have singular, or resonant points at $r=r_A$, where the phase speed of the wave perturbation coincides with that of relativistic Alfv\'en waves. 
Like the corotation singularity, this singularity is also physical, and results from the resonant interaction of  perturbations with relativistic Alfv\'en waves.

\citet{Pariev96} have discussed the proper way to handle these singularities, however, since singular points 
are found only for real values of $\omega$ and since we are interested in unstable modes and therefore complex 
values of $\omega$, we can assume that our integration will always avoid singular points.

\end{enumerate}

%
%

\section{Results}
%
%

Before discussing in detail our results, however, we recall in section \ref{subsec:parameters} the units we will use, the full set of parameters  defining our problem and which ones will be investigated in more detail in the present work.
We then preliminarly examine the static case in subsection \ref{subsec:static}, the full case in subsection \ref{subsec:moving} and finally a more detailed analysis of the CDI is performed in subsection \ref{subsec:cd}.

\subsection{Parameters and units}
\label{subsec:parameters}
%
%

As discussed in Section \ref{sec:equilibrium},  the basic equilibrium state is determined by five parameters: the jet density contrast $\eta$, the bulk flow velocity which is derived from $\gamma_c$, the rotation parameter $\alpha$, the pitch on the axis $P_c$ and the ratio $M^2_a$ between the energy density of the matter and the magnetic energy density. 
We notice that our parameter $M_a^2$ is the inverse of the  magnetization parameter often used in the literature. 
In the present paper, as already discussed, we consider only the case without rotation, so $\alpha$ is kept fixed to $0$, and we also consider jets with density equal to that of the ambient medium, so $\eta$ is kept fixed to $1$. 
We focus our analysis on  the dependence on the three parameters $\gamma_c$, $P_c$ and $M_a$.  
In the following discussion, instead of $M_a$ we will make use of the quantity  $M_a v_c$ which, in the Newtonian limit, corresponds to the ratio  between the jet speed and the Alfv\'en speed, while in the relativistic limit reduces to $M_a$. 
The unit of velocity is the light speed $c$ and we choose as unit of length the jet radius $r_j$, so, unless otherwise specified, the growth rate is expressed in units of $c/r_j$.   
An additional parameter is represented by the width $a$ of the current distribution, this parameter, unless otherwise specified, is kept fixed to $a = 0.6$. 
Of course, in the static case, the jet radius has no significance and is an arbitrary measure, however we keep it as the unit of length for consistency with the general case. 

\subsection{The static case, $v_z=0$.}
\label{subsec:static}
%
%

Static columns in the Newtonian case have been already studied by many authors \citep[see e.g.][]{Appl00, Bonanno08, Bonanno11} and differ one another in many respects, but mainly for the parameter range considered and for the equilibrium magnetic field configuration. 
In particular the studies by \citet{Bonanno08} consider cases with subthermal magnetic field strength. 
In the present work, conversely, we neglect the pressure term and our regime of investigation addresses the case of suprathermal field strengths and can be compared with that of \citet{Appl00}, that is performed in the same regime. 

In the study of CDI, a very important role is played by resonant surfaces, i.e. the surfaces where the condition $\vec{k} \cdot \vec{B} = 0$ is satisfied. 
On these surfaces, in fact, the stabilizing effect of magnetic tension is absent and these regions are therefore more prone to instability. The resonant condition can be rewritten as 
\begin{equation}\label{eq:resonance}
 \vec{k} \cdot \vec{B} =  k B_z + \frac{m}{r} B_\phi = k P + m = 0 \,.
\end{equation}
This condition can be verified only for positive $m$ and, for a constant pitch distribution like that used by \citet{Appl00}, marks the stability boundary:  wavenumbers larger than this critical limit are stable, while smaller wavenumbers are unstable. 
\begin{figure}
   \centering
      \includegraphics[width=9cm]{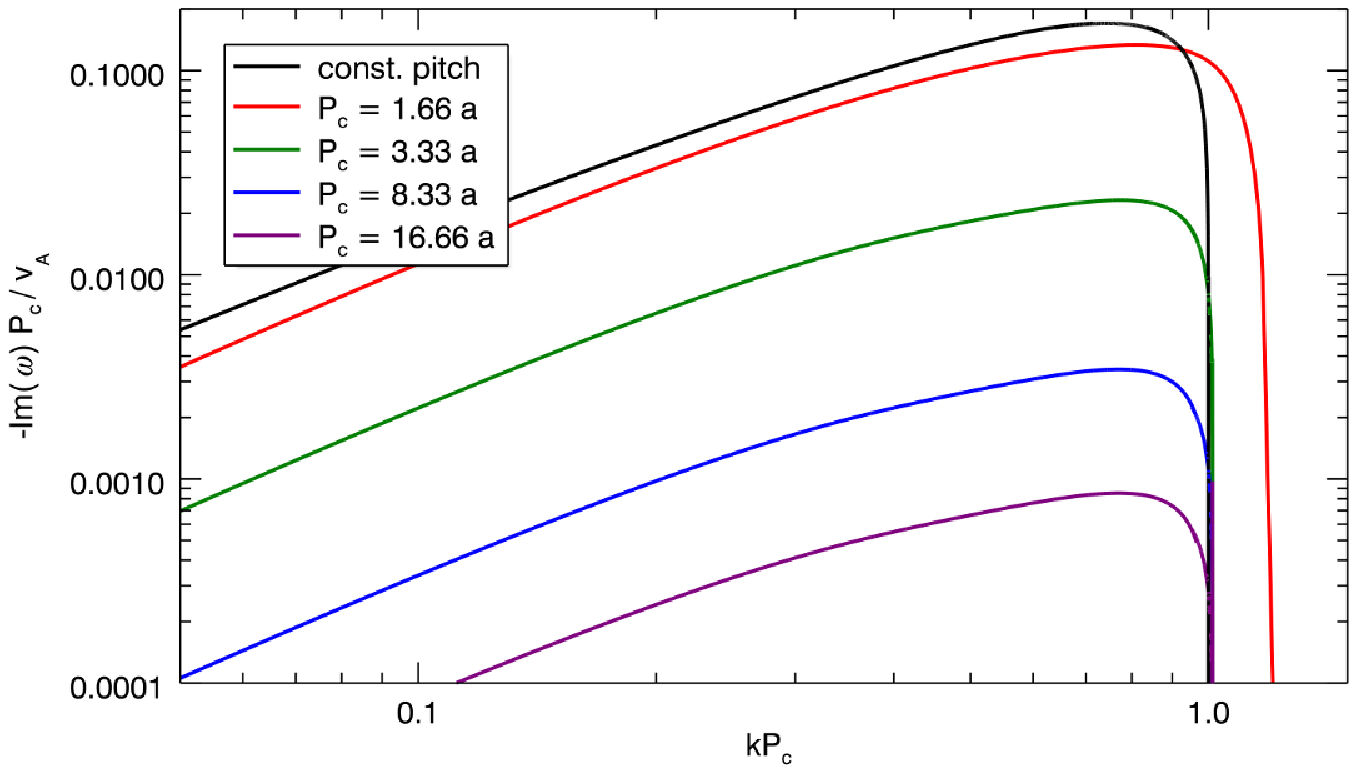} 
   \includegraphics[width=9cm]{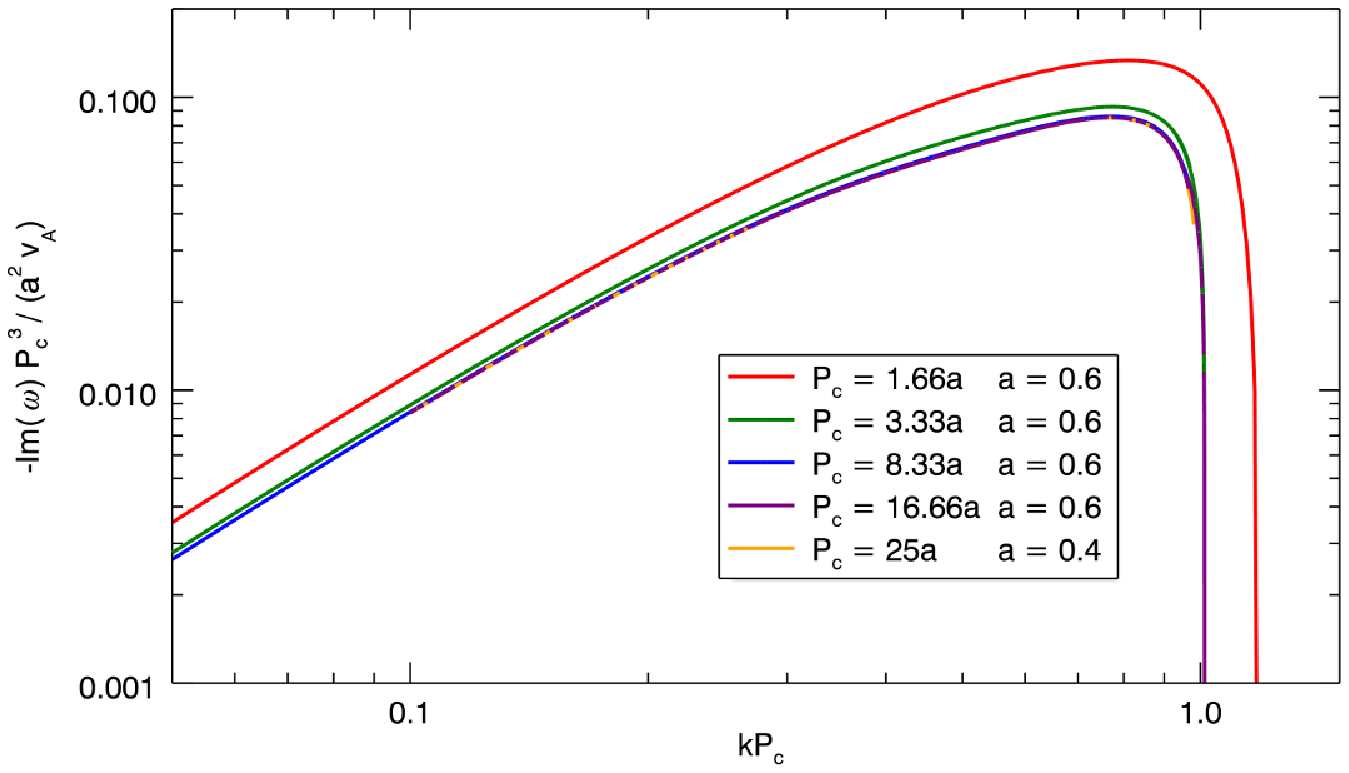} 
   \caption{\small Normalized plots of the growth rate as a function of the wavenumber for a static column and for different values of the pitch parameter $P_c$. 
   The values of $P_c$ for the different curves are reported in the legend and, for comparison, in the top panel, we plot also the case of a configuration with constant pitch.   
   In the bottom panel we show how well the scaling given by Eq. (\ref{eq:scaling}) reproduces our results, a significant deviation can be observed only for $P_c/a = 1.66$. }
   \label{fig:staticmhd}
\end{figure}
This behavior can be observed in the top panel of Fig. \ref{fig:staticmhd}, where we plot the growth rate as a function of the wavenumber for the case with $m = 1$ for the constant pitch configuration (black curve, top panel) and for our equilibrium model (colored lines).
In the top panel we measure the wavenumber in units of $1/P_c$ and the growth rate in units of $v_A/P_c$. 
In these units, the constant pitch solutions of \cite{Appl00} are all represented by the single black curve independently of the pitch value.
Instability is  present for all wavenumbers $k < 1/P_c$  (i.e. for wavenumbers lower than the threshold given by the resonance condition Eq. \ref{eq:resonance}) and, for $k \rightarrow 0$, the growth rate decreases as a power law.   
We note that, while we employ radiation condition at infinity, \citet{Appl00} in most of their calculations made use of boundary conditions consistent with a rigid wall set at a finite radial distance. 
This condition introduces a new scale and restricts the unstable range in wavenumbers, with a small wavenumber cutoff.  

\begin{figure}
   \centering
   \includegraphics[width=9cm]{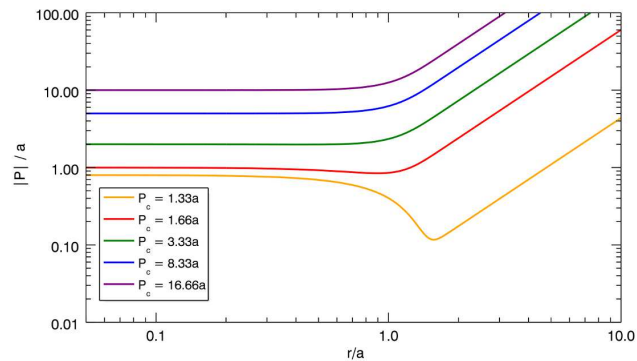} 
   \caption{\small Plot of the pitch profile as a function of $r/a$.  
   The different curves refer to different values of the parameter $P_c/a$, whose values are reported in the legend.}
   \label{fig:pitch}
\end{figure}
Our equilibria are characterized by two length scales:  the pitch value on the axis $P_c$ and the width of the current distribution $a$ and therefore we have the additional parameter $P_c/a$.  
The pitch profile $P(r)$ plays a fundamental role for the stability properties and, for our equilibrium configurations, it is shown in Fig. \ref{fig:pitch} for different values of $P_c/a$. 
It is always characterized by an increase at large radii, owing to the confinment of the current inside $r < a$, while  in the central region we can distinguish two regimes: for $P_c/a \gg 1$  the pitch is  constant up to $r/a = 1$ while for  $P_c/a \sim 1$ it decreases to a minimum value at $r/a \sim 1.5$ immediately after the inner flat region.
Furthermore, there exists a critical value of  $P_c/a = 1.33$  below which equilibrium configuration are not possible. 
Below this critical value, the equilibrium condition would require a negative value of $B_z^2$ meaning that the longitudinal field pressure gradient is no longer able to balance the inward force of the azimuthal magnetic field. 
The behavior of the growth rate in the first regime ($P_c/a \gg 1$) has a form similar to the constant pitch situation, but its  value decreases as the parameter $P_c/a$ increases, as it is shown in the top panel of Fig. \ref{fig:staticmhd}. 
In the same limit the growth rate takes the asymptotic form 
\begin{equation}\label{eq:scaling}
 \hbox{Im} (\omega) \sim  \frac{v_A}{P_c} \left( \frac{a}{P_c} \right)^2 f(k P_c)\,,
\end{equation}
where the function $f(kP_c)$ is independent from $P_c/a$. 
The validity of the previous scaling law for any value of $P_c/a$ is demonstrated  in the bottom panel of Fig. \ref{fig:staticmhd}, where we plot $\hbox{Im}(\omega) P_c^3 / a^2 v_A$ as a function of $k P_c$.  
The curves in the figures, therefore, represent the function $f(kP_c)$ and should be independent from $P_c/a$. 
In fact, the purple ($P_c/a = 16.66$, $a=0.6$), blue ($P_c/a = 8.33$, $a=0.6$) and orange ($P_c/a = 25$, $a=0.4$) curves are overimposed and almost coincident with the green curve corresponding to $P_c/a = 2.66$, $a = 0.6$.  
A significant deviation from the above scaling is observed only for $P_c/a = 1.66$, as shown by the red curve.  
In this case, as discussed above, we have a region of decreasing pitch whose effect is to widen the instability range to larger values of $k P_c$ thus increasing the growth rate above the value predicted by the scaling law and moving the maximum towards somewhat larger values of $k P_c$. 

For a static configuration, relativistic effects are introduced only by increasing the magnetic field strength and the previous results shown in Fig. \ref{fig:staticmhd} remain the same provided the Alfv\'en velocity  is replaced by the correct relativistic expression
\begin{equation}
 v_A = \frac{|\vec{B}|}{\sqrt{w + \vec{B}^2}}\,,
\end{equation}
where $w$ is the gas enthalpy.

\subsection{The case with $v_z \neq 0$.}
\label{subsec:moving}
%
%

\begin{figure}
   \centering
  \includegraphics[width=15cm]{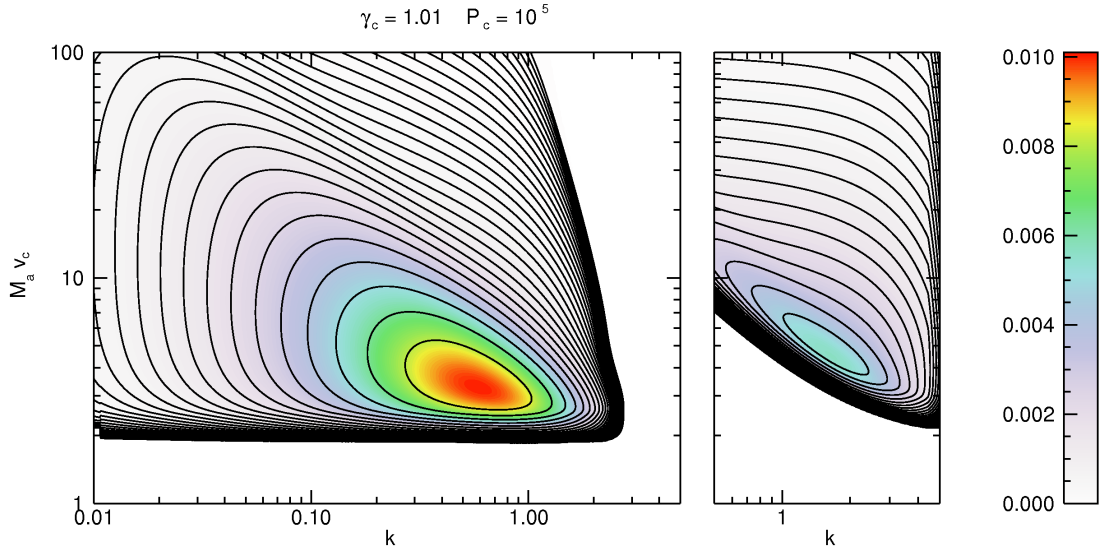} 
   \caption{\small Distribution of the growth rate as function of the wavenumber and of $M_a v_c$ for the case with $\gamma_c = 1.01$ and $P_c = 10^5$. As discussed in the text, the results for the two cases $m = 1$ and $m = -1$ coincide. In the left panel we have the Kelvin-Helmholtz ordinary mode, while  in the right panel we have the first reflected Kelvin-Helmholtz mode. The levels are equispaced in logarithmic scale from $10^{-5}$  to the maximum value of the growth rate.}
   \label{fig:P10000MHD}
\end{figure}

\begin{figure}
   \centering
  \includegraphics[width=15cm]{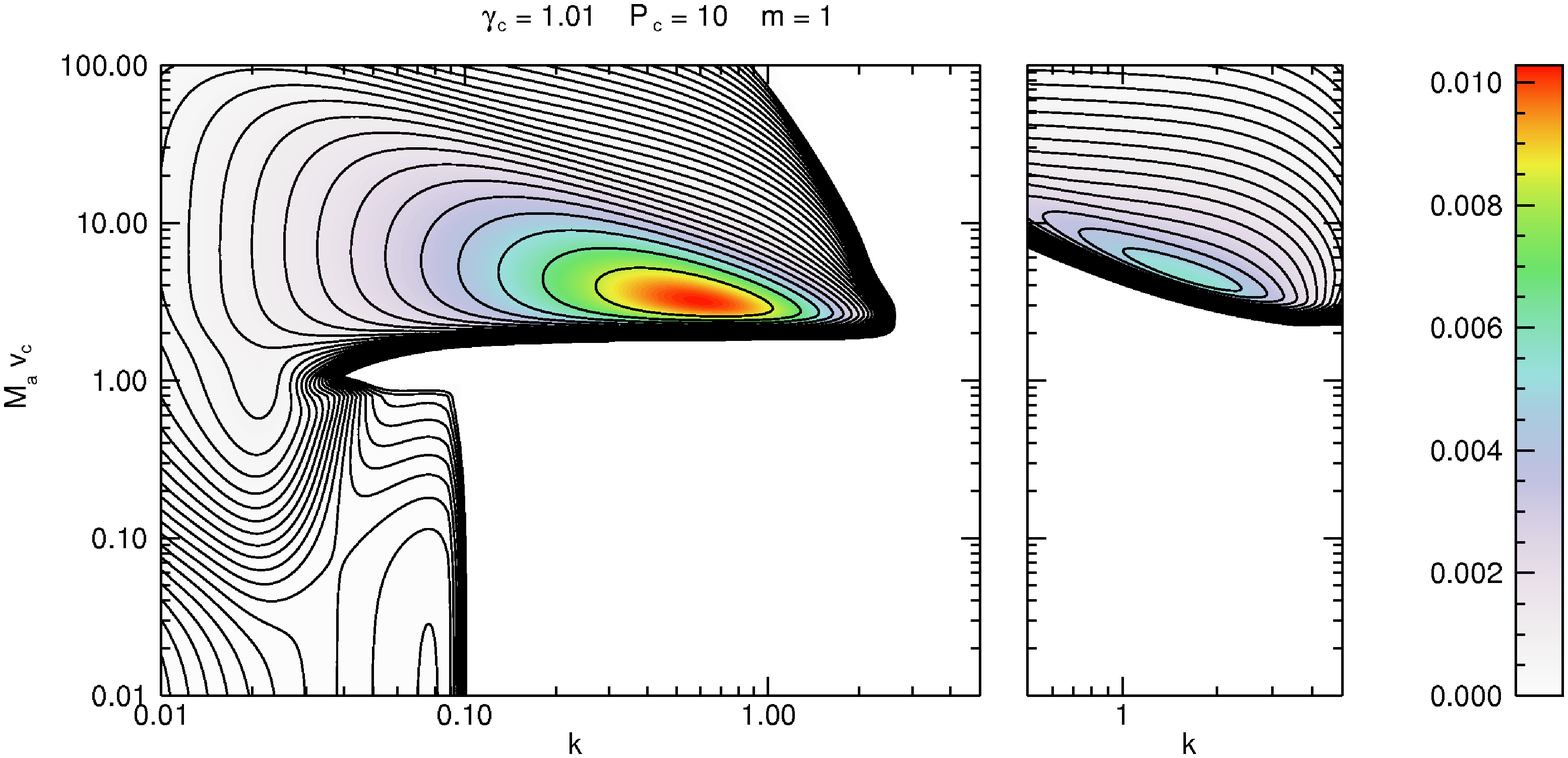} 
    \includegraphics[width=15cm]{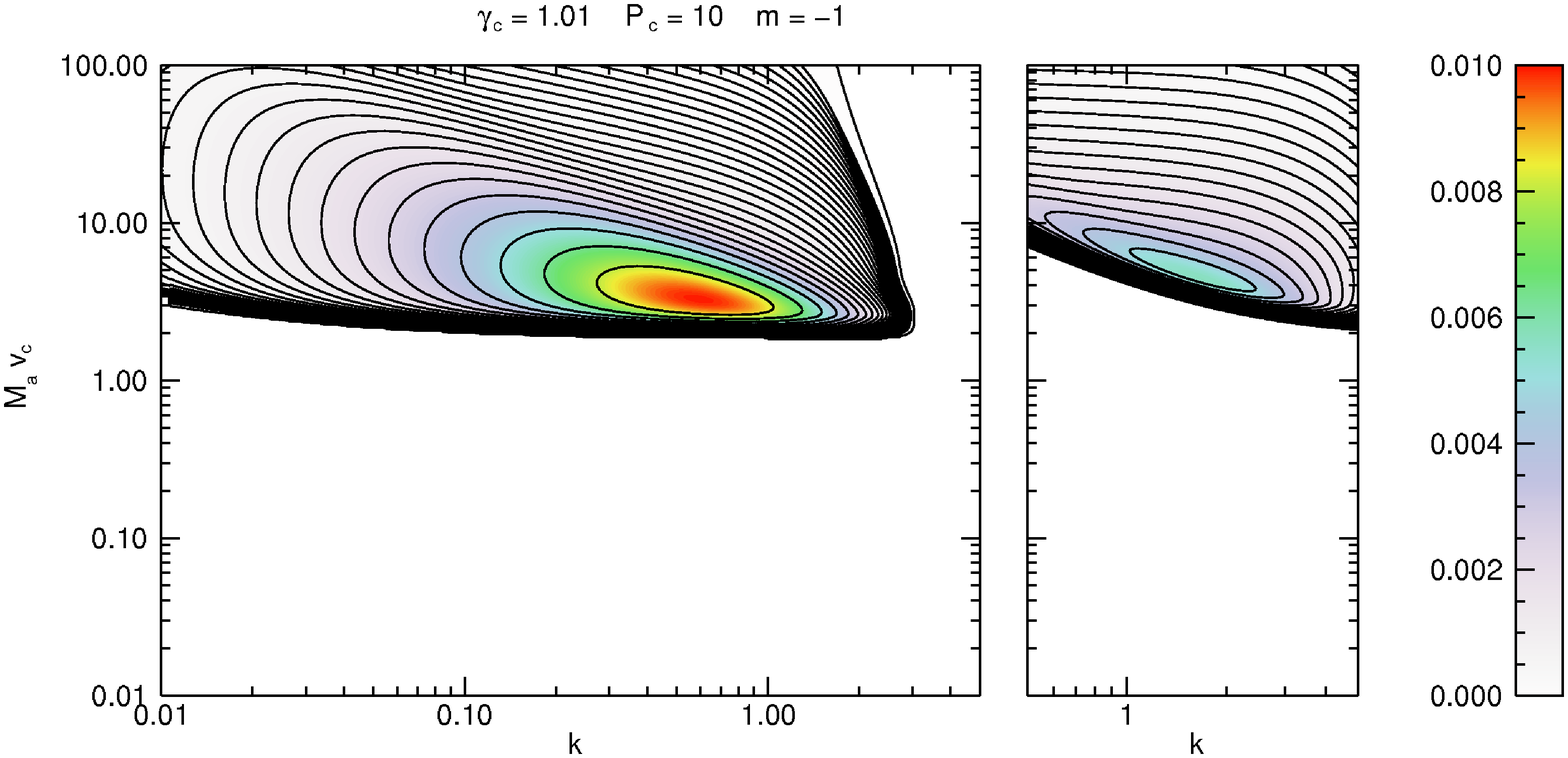} 
   \caption{\small Distribution of the growth rate as function of the wavenumber and of $M_a v_c$ for the case with $\gamma_c = 1.01$ and $P_c = 10$.  The top panels refer to $m = 1$ while the bottom panels refer to $m = -1$. In the left panels, for high values of $M_a v_c$, we have the ordinary mode of the Kelvin-Herlmholtz instability, for low values of $M_a v_c$ the case $m =1$ (top) shows the current driven instability, while the case $m=-1$ (bottom) is stable.  In the  right panels we have the first reflected Kelvin-Helmholtz mode.   The levels are equispaced in logarithmic scale from $10^{-5}$  to the maximum value of the growth rate.
   }
   \label{fig:P10MHD}
\end{figure}

\begin{figure}
   \centering
   \includegraphics[width=15cm]{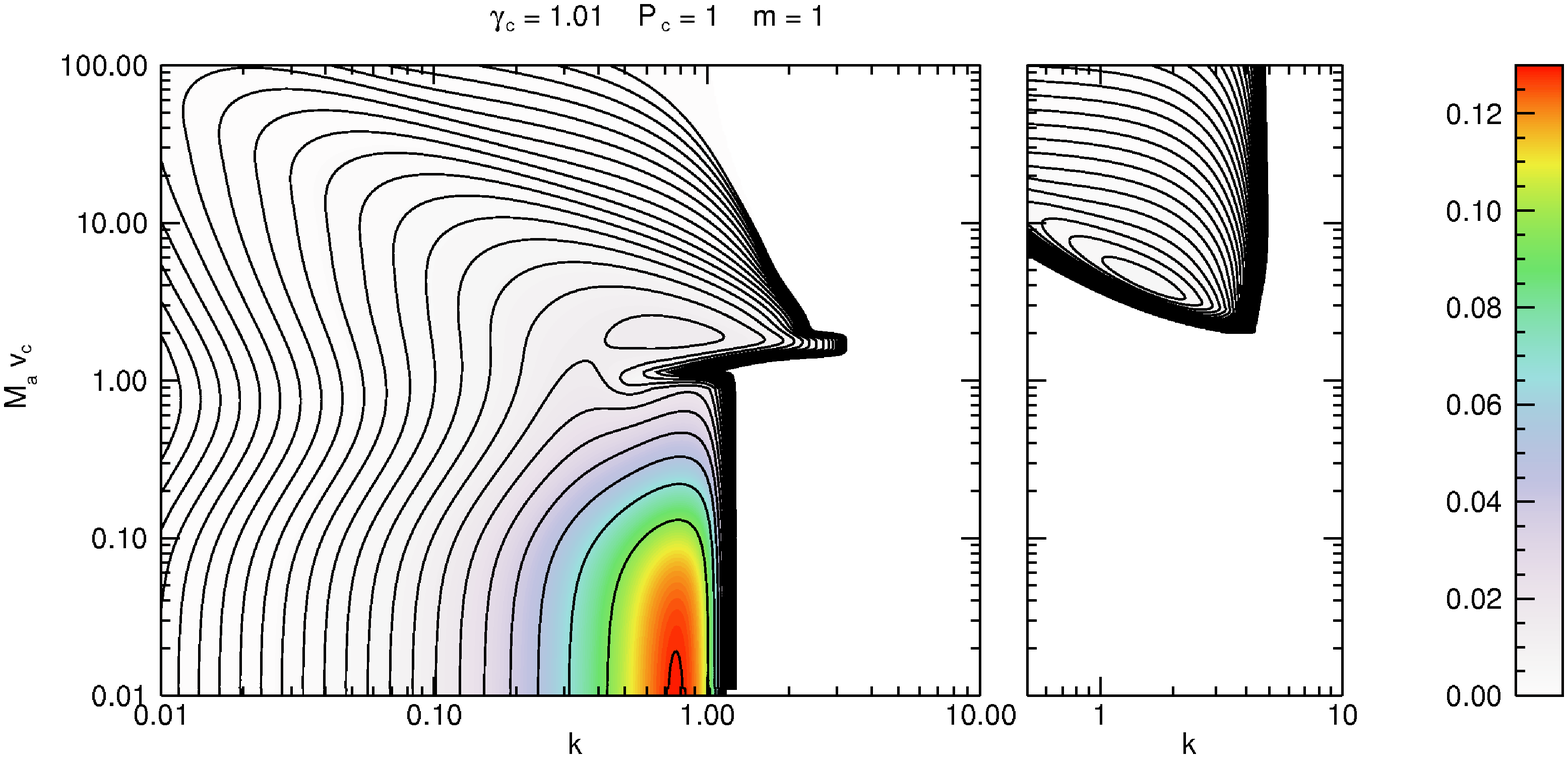} 
   \includegraphics[width=15cm]{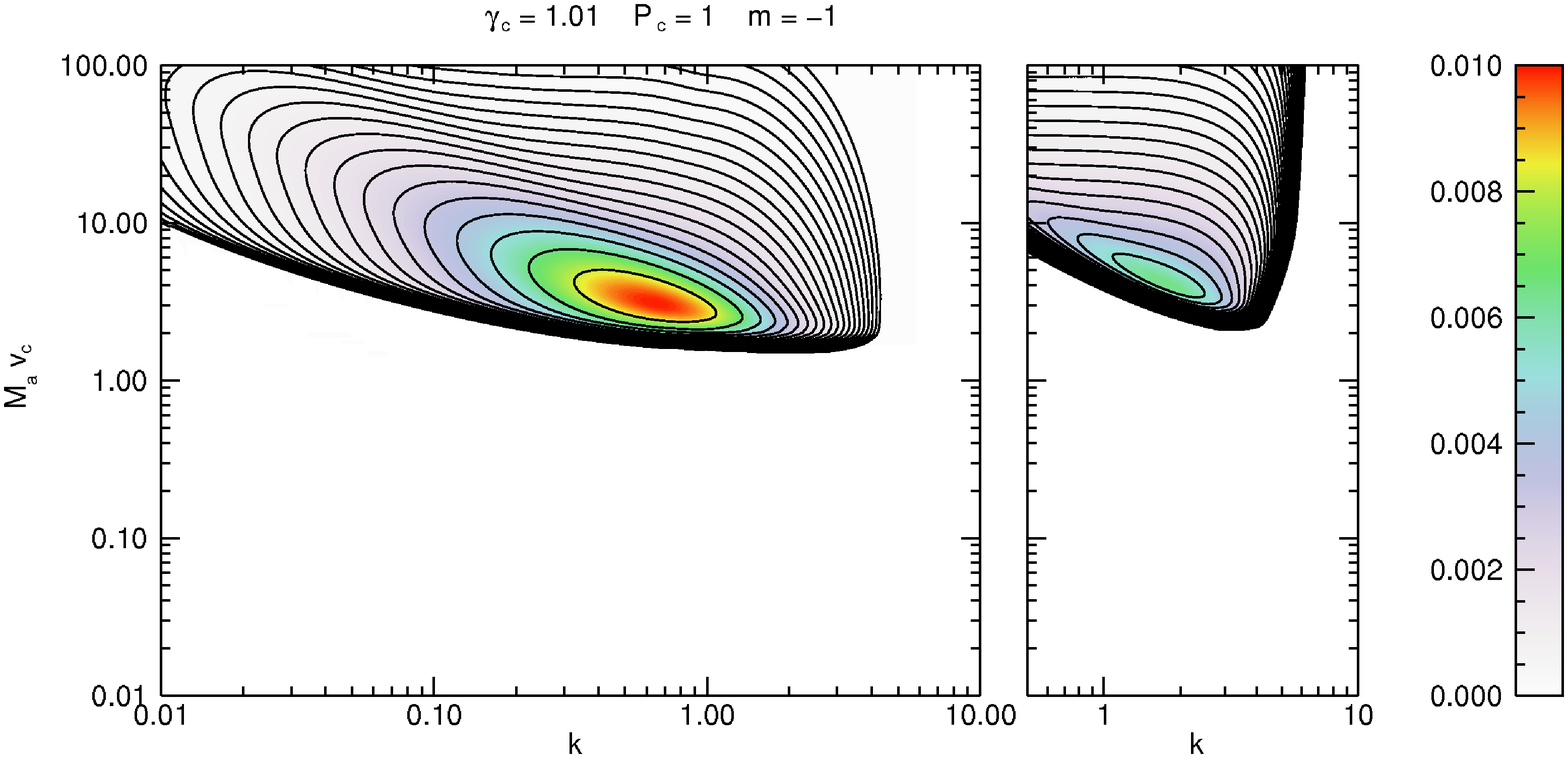} 
   \caption{\small Distribution of the growth rate as function of the wavenumber and of $M_a v_c$ for the case with $\gamma_c = 1.01$ and $P_c = 1$. 
The top panels refer to $m = 1$ while the bottom panels refer to $m = -1$. In the left panels, for high values of $M_a v_c$, we have the ordinary mode of the Kelvin-Herlmholtz instability, for low values of $M_a v_c$ the case $m =1$ (top) shows the current driven instability, while the case $m=-1$ (bottom) is stable.  
In the  right panels we have the first reflected Kelvin-Helmholtz mode.   
The levels are equispaced in logarithmic scale from $10^{-5}$ to the maximum value of the growth rate. }
   \label{fig:P1MHD}
\end{figure}

We start the discussion of the dynamic case considering a flow moving at $\gamma_c = 1.01$, corresponding to a non-relativistic $v_c \sim 0.14$. 
In this case the results obtained in the Newtonian limit coincide almost exactly with those obtained with the full relativistic treatment. 
An overview of the mode structure can be gained by looking at Figs. \ref{fig:P10000MHD}, \ref{fig:P10MHD}, \ref{fig:P1MHD}, where we show the behavior of the growth rate as a function of the wavenumber $k$ and of $M_a v_c$ for $P_c = 10^5$,  $P_c = 10$ and $P_c = 1$, respectively.
Figs. \ref{fig:P10MHD} and \ref{fig:P1MHD} present the results for $m=1$ in the upper panels and for $m=-1$ in the lower panels, while in Fig. \ref{fig:P10000MHD} the system has no way to distinguish between $m=1$ and $m=-1$ and the results for the two modes are coincident.  
As mentioned, we now expect the appearence of two types of instability, namely, Kelvin-Helmholtz instabilities (KHI) and current driven instabilities (CDI). 
Moreover, in the case of a purely longitudinal field, CDI are absent and only KHI may be effective.  
This is the case of Fig. \ref{fig:P10000MHD}, where we show the results for $P_c = 10^5$, i.e. for a magnetic field almost exactly longitudinal.  
The two panels in the figure refer to the ordinary mode and to the first reflected mode \citep[for a discussion on reflected modes, see e.g.][]{Bodo89}. 
From the figure, we can see that  the jet is stable below $M_a v_c \sim v_c/v_A \sim 2$, the maximum growth rate for the ordinary mode is $\hbox{Im}(\omega_{\max})\sim 0.01$ and it is found at $k \sim 0.6$ and $M_a v_c \sim v_c/v_A \sim 3$. 
Moving towards higher values of $v/v_A$, the relative maximum shifts towards smaller  values of $k$ and our results show that both the wavenumber of the maximum and the maximum itself scale as $v_A/v_c$.
The first reflected mode (right panel) has a smaller growth rate, $\hbox{Im}(\omega_{\max}) \sim 0.005$, and remains unstable at larger values of $k$ as expected.

Fig. \ref{fig:P10MHD} refers to $P_c = 10$ with the upper and lower panels showing, respectively, the growth rates computed for the $m=1$ and $m=-1$ modes.
Clearly, the upper left panel shows a region of instability for small values of $M_a v_c$ which is completely absent in the lower panel.
This instability region corresponds to onset of CDI modes that become stable for $m=-1$ since the resonance condition (Eq. \ref{eq:resonance}) cannot be satisfied. 
For small values of $M_a v_c$, the instability behavior is very similar to the static case discussed in the previous subsection and its growth rate is two order of magnitude smaller than that of the KHI. 
The mode becomes stable for $k > 0.1$, which corresponds to the stability limit given by the condition $k P_c = 1$. 
For larger values of $M_a v_c$ the CD mode merge with the KH mode and shows essentially no difference with respect to the previous case.

The results for $P_c = 1$ are shown in Fig. \ref{fig:P1MHD}. 
As expected, the CDI moves towards higher values of $k$ as the stability limit $k P_c = 1$ gives a limiting value of $k = 1$. 
The growth rate of the CD mode also increases and scales as $1/P_c^3$ as in the static case eventually becoming dominant over the KHI. 
For $m = -1$ (lower panels) the CDI is absent and the KHI presents only slight differences with the cases at larger values of $P_c$.

\begin{figure}
   \centering
  \includegraphics[width=15cm]{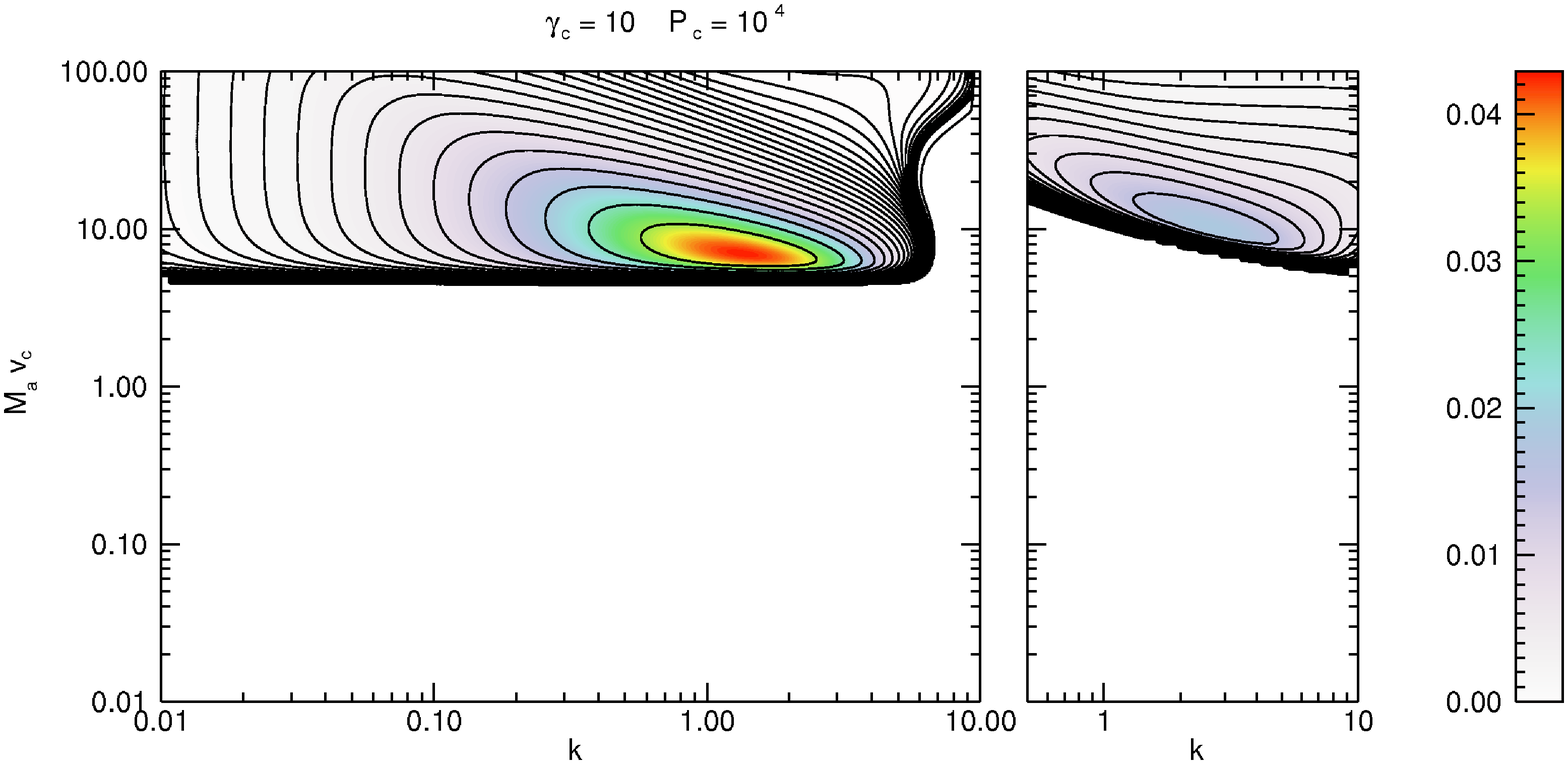} 
   \caption{\small Distribution of the growth rate as function of the wavenumber and of $M_a v_c$ for the case with $\gamma_c = 10$ and $P_c = 10^4$ (correponding to $P_c = 10^5$ in the rest frame of the jet). As discussed in the text, the results for the two cases $m = 1$ and $m = -1$ coincide. In the left panel we have the Kelvin-Helmholtz ordinary mode, while  in the right panel we have the first reflected Kelvin-Helmholtz mode.   The levels are equispaced in logarithmic scale from $10^{-5}$ to the maximum value of the growth rate. 
   }
   \label{fig:P10000RMHD}
\end{figure}

\begin{figure}
   \centering
    \includegraphics[width=15cm]{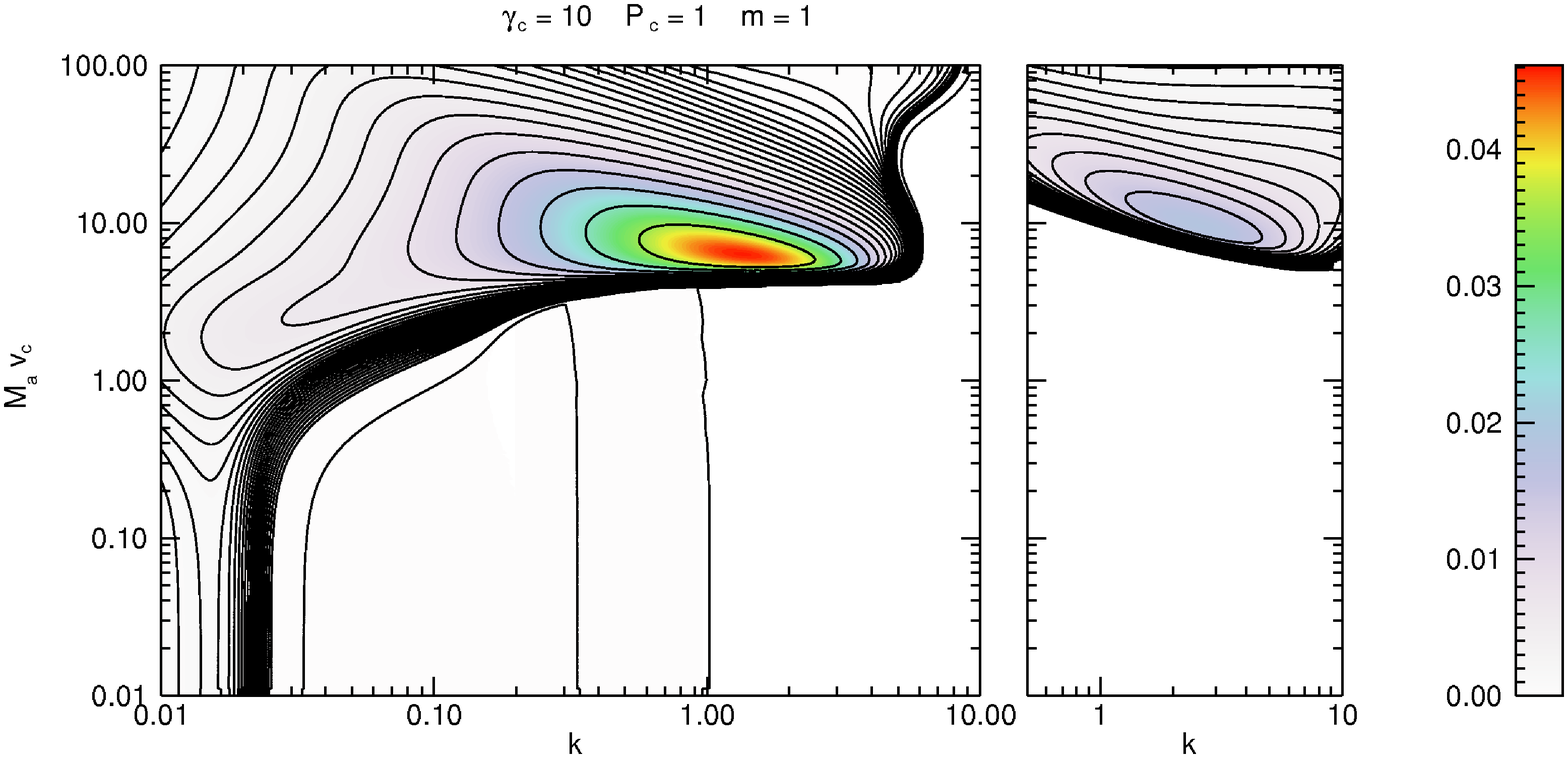} 
    \includegraphics[width=15cm]{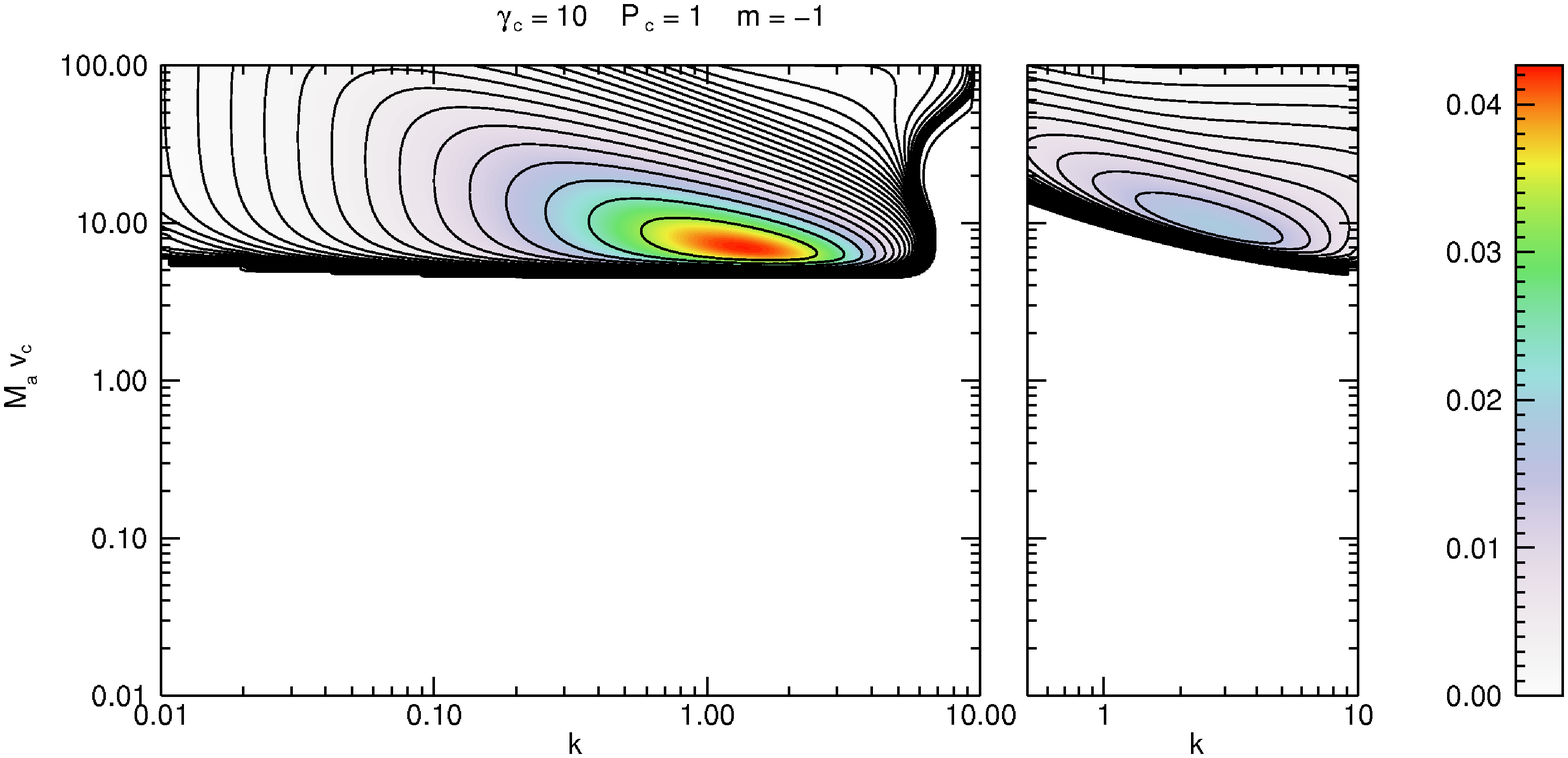} 
  \caption{\small Distribution of the growth rate as function of the wavenumber and of $M_a v_c$ for the case with $\gamma_c = 10$ and $P_c = 1$ (correponding to $P_c = 10$ in the rest frame of the jet).  The top panels refer to $m = 1$ while the bottom panels refer to $m = -1$. In the left panels, for high values of $M_a v_c$, we have the ordinary mode of the Kelvin-Herlmholtz instability, for low values of $M_a v_c$ the case $m =1$ (top) shows the current driven instability which is splitted in two branches (note that the region between $k=0.04$ and $k=0.3$ is stable), while the case $m=-1$ (bottom) is stable.  In the  right panels we have the first reflected Kelvin-Helmholtz mode.  The levels are equispaced in logarithmic scale from $10^{-5}$ to the maximum value of the growth rate. 
}
   \label{fig:P10RMHD}
\end{figure}

\begin{figure}
   \centering
   \includegraphics[width=15cm]{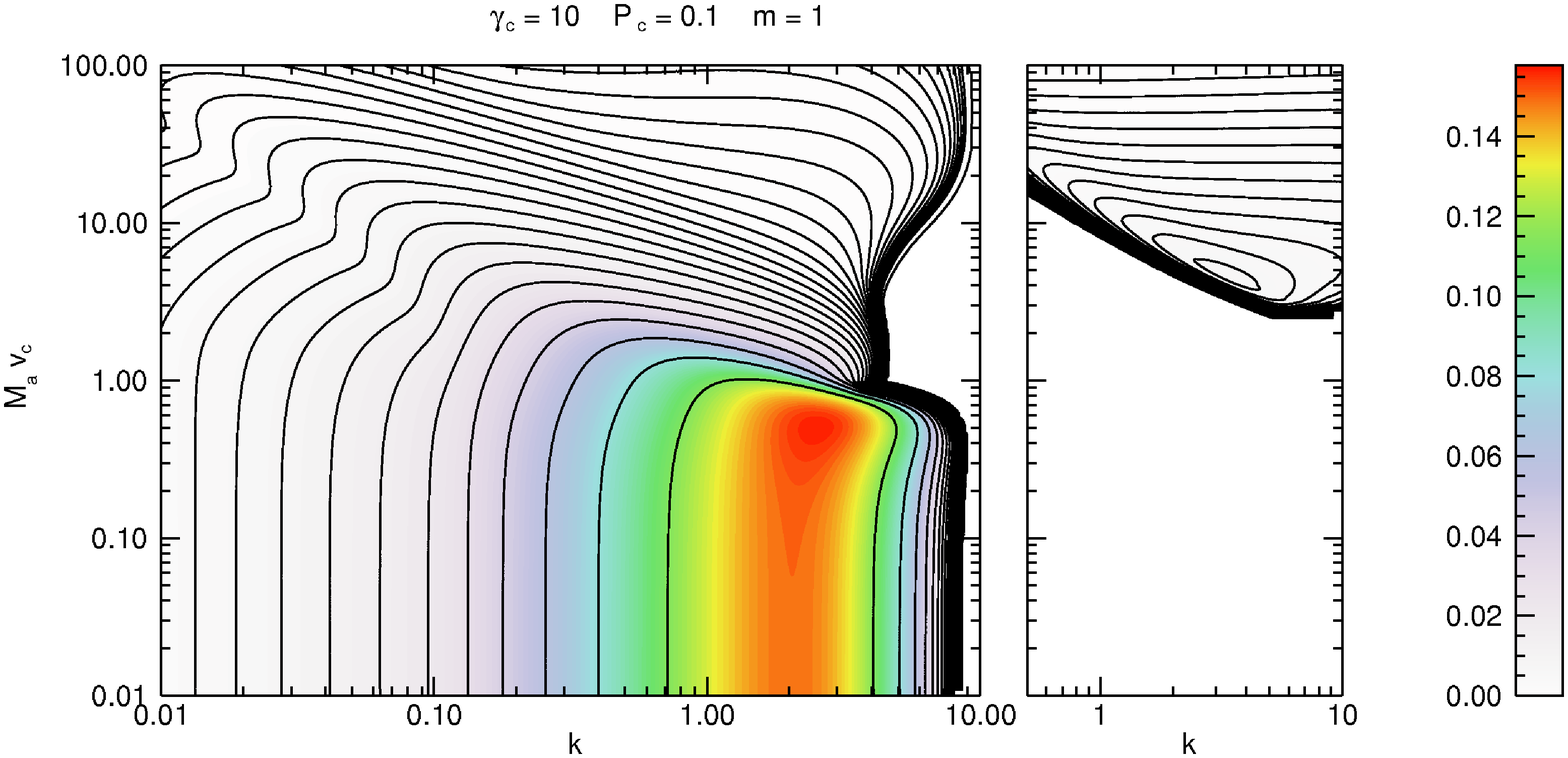} 
   \includegraphics[width=15cm]{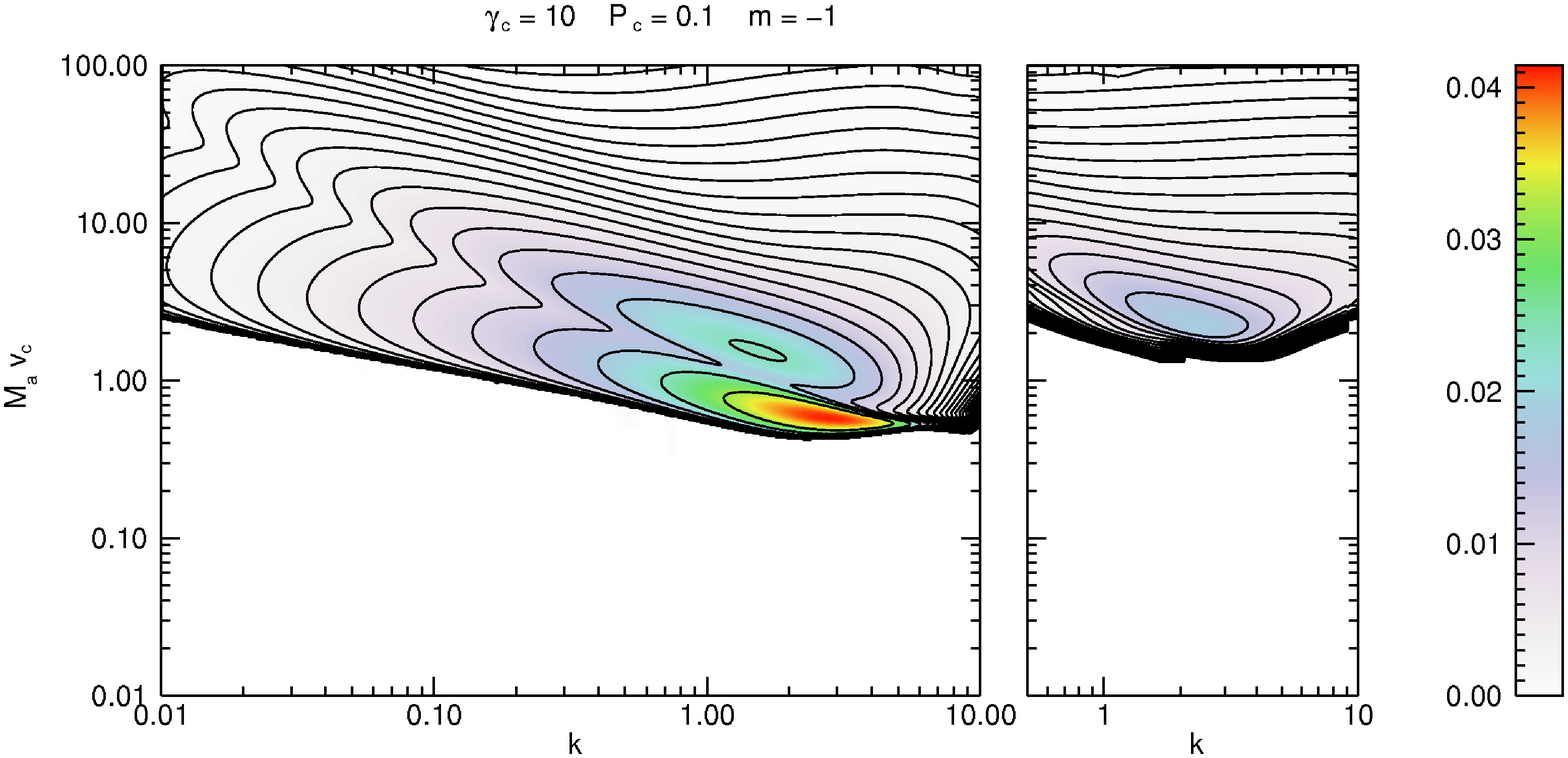} 
   \caption{\small Distribution of the growth rate as function of the wavenumber and of $M_a v_c$ for the case with $\gamma_c = 10$ and $P_c = 0.1$  (correponding to $P_c = 1$ in the rest frame of the jet) . The top panels refer to $m = 1$ while the bottom panels refer to $m = -1$. As discussed in the text, for $m = 1$ (top), the Kelvin-Helmholtz ordinary mode and the current driven mode merged. in the right panel we have the first reflected Kelvin-Helmholtz mode.   The levels are equispaced in logarithmic scale from $10^{-5}$ to the maximum value of the growth rate.  }
   \label{fig:P1RMHD}
\end{figure}
Increasing the flow velocity up to Lorentz $\gamma_c = 10$, we show the growth rate behavior as a function of $k$ and $M_a v_c$ in Figs  \ref{fig:P10000RMHD}, \ref{fig:P10RMHD}, \ref{fig:P1RMHD} corresponding, respectively, to $P_c = 10^4$, $P_c = 1$ and $P_c = 0.1$.  
We remark that the pitch in the comoving frame is obtained by multiplying the pitch in the lab frame by $\gamma$ and therefore the chosen values of $P_c$ in the comoving frame are exactly the same ones used for the the classical case. 
The upper and lower panels in Figs. \ref{fig:P10RMHD} and \ref{fig:P1RMHD} refer respectively, to $m=1$ and $m=-1$.
Similarly to the classical case, we do not have any difference between the $m=1$ and $m=-1$ modes for $P_c=10^4$, see Fig.  \ref{fig:P10000RMHD}.  
The right panels in each figure show, as before,  the reflected KH mode. 
The general instability behavior is quite similar to the classical case and we can easily recognize the KHI region and the CDI region. 
Focusing on the KHI, we see that the stability boundary and, consequently, the position of the maximum growth rate has moved towards larger values of $M_a$ (see Figs. \ref{fig:P10000RMHD}, \ref{fig:P10RMHD} left and right panels) as it is expected to happen for relativistic flows \citep{Osmanov08}.  
Decreasing $P_c$, the KH instability boundary moves towards smaller values of $M_a v_c$, as it is evident in the lower panels of Fig. \ref{fig:P1RMHD} corresponding to  $m=-1$ and in the right upper panel of the same figure (reflected mode $m=1$).  
Indeed, the stabilizing longitudinal component of magnetic field decreases with decreasing $P_c$.  
In the upper left panel of Fig. \ref{fig:P1RMHD} we see that, for $m=1$, the KHI and CDI have merged.  
Moreover, for each value of $P_c$, an increase of $M_a v_c$ leads to a shift of the relative maximum of the growth rate towards smaller values of $k$ (see Figs. \ref{fig:P10000RMHD}, \ref{fig:P10RMHD}, \ref{fig:P1RMHD}) and both the wavenumber of the maximum and the maximum itself scale as $1/M_a v_c$ precisely as in the Newtonian limit.  

\begin{figure}
   \centering
  \includegraphics[width=9cm]{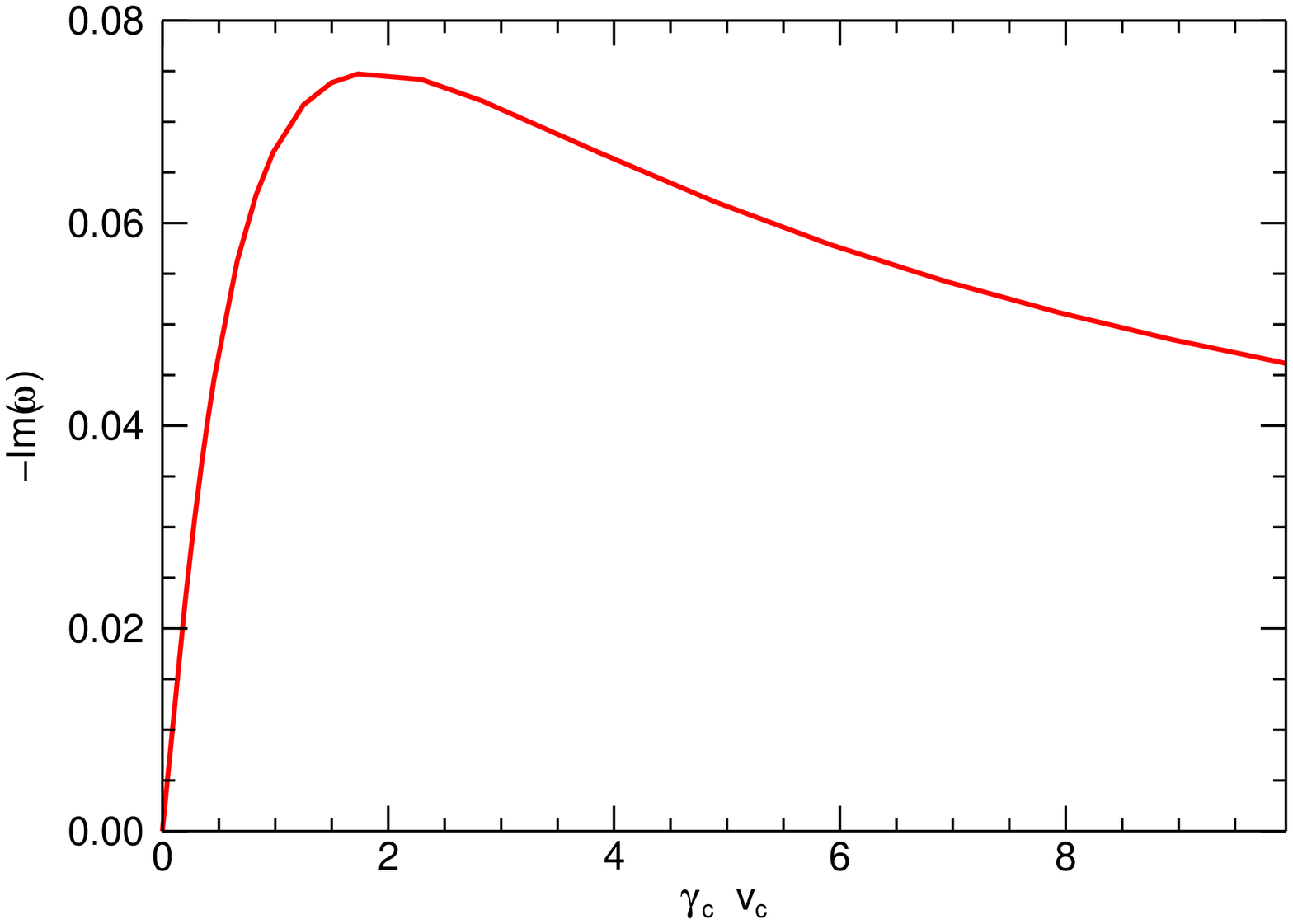} 
     \caption{\small Plot of the maximum growth rate of the Kelvin-Helmholtz instability as a function of $\gamma_c v_c$ for $P_c = 1$ and $m = 1$.}
   \label{fig:maxkh}
\end{figure}
In Fig. \ref{fig:maxkh}  we plot  the value of the maximum growth rate of the KH mode as a function of $\gamma_c v_c$ for $P_c = 1$ and $m = 1$.  
This curve can be considered representative for any value of the pitch since, as we have seen, the growth rate of the KHI is only weakly dependent on this parameter.  
In the Newtonian limit, for small values of $\gamma_c v_c$, the growth rate becomes essentially proportional to $v_c$, reaches a maximum at $\gamma_c v_c \sim 2$ and then progressively decreases.
This result is in agreement with the work of \cite{Osmanov08} that found that relativistic motion plays a stabilizing role on the growth of the KHI.

\subsection{The current driven mode}
\label{subsec:cd}
%
%

\begin{figure}
  \centering
  \includegraphics[width=9cm]{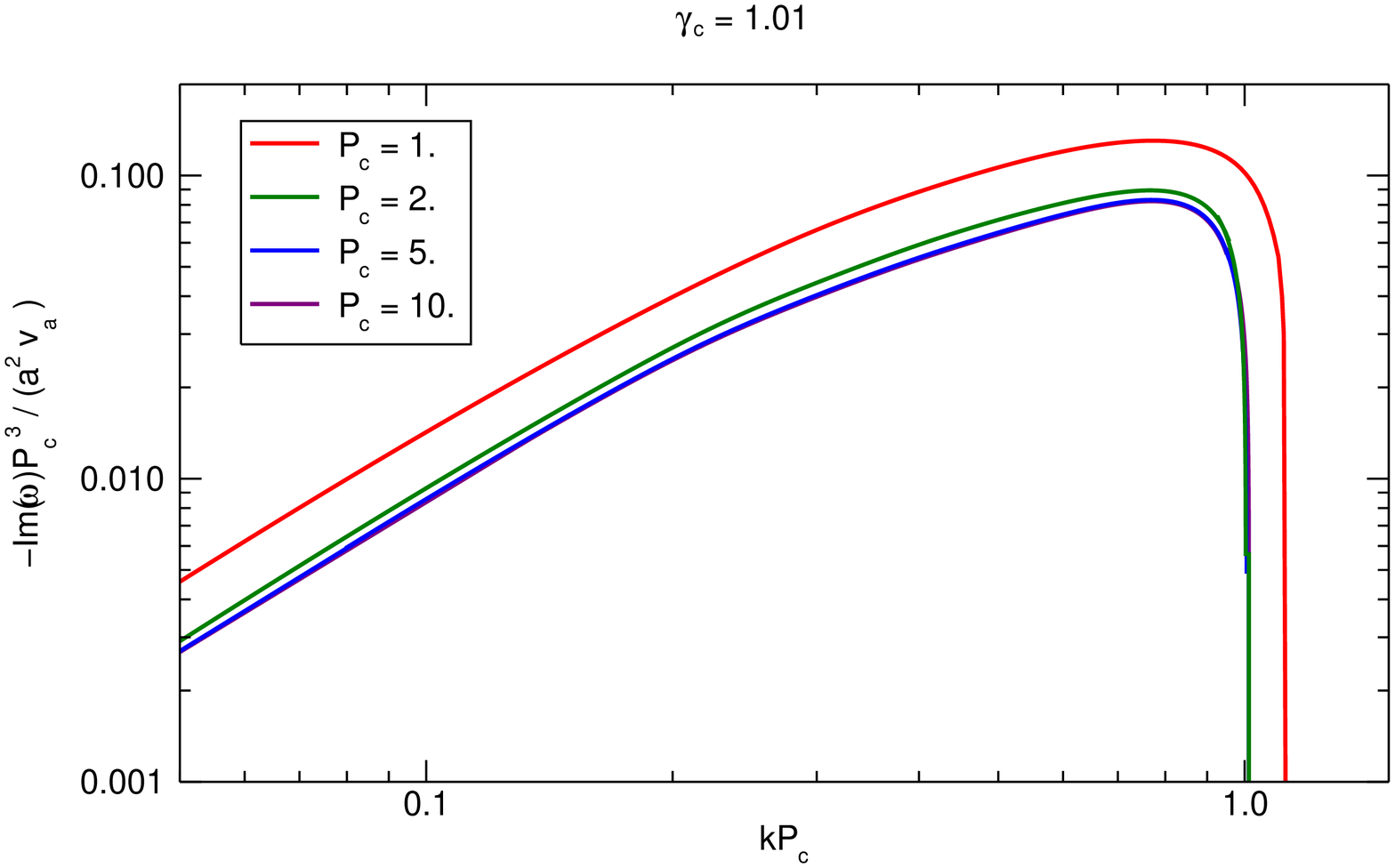} 
  \includegraphics[width=9cm]{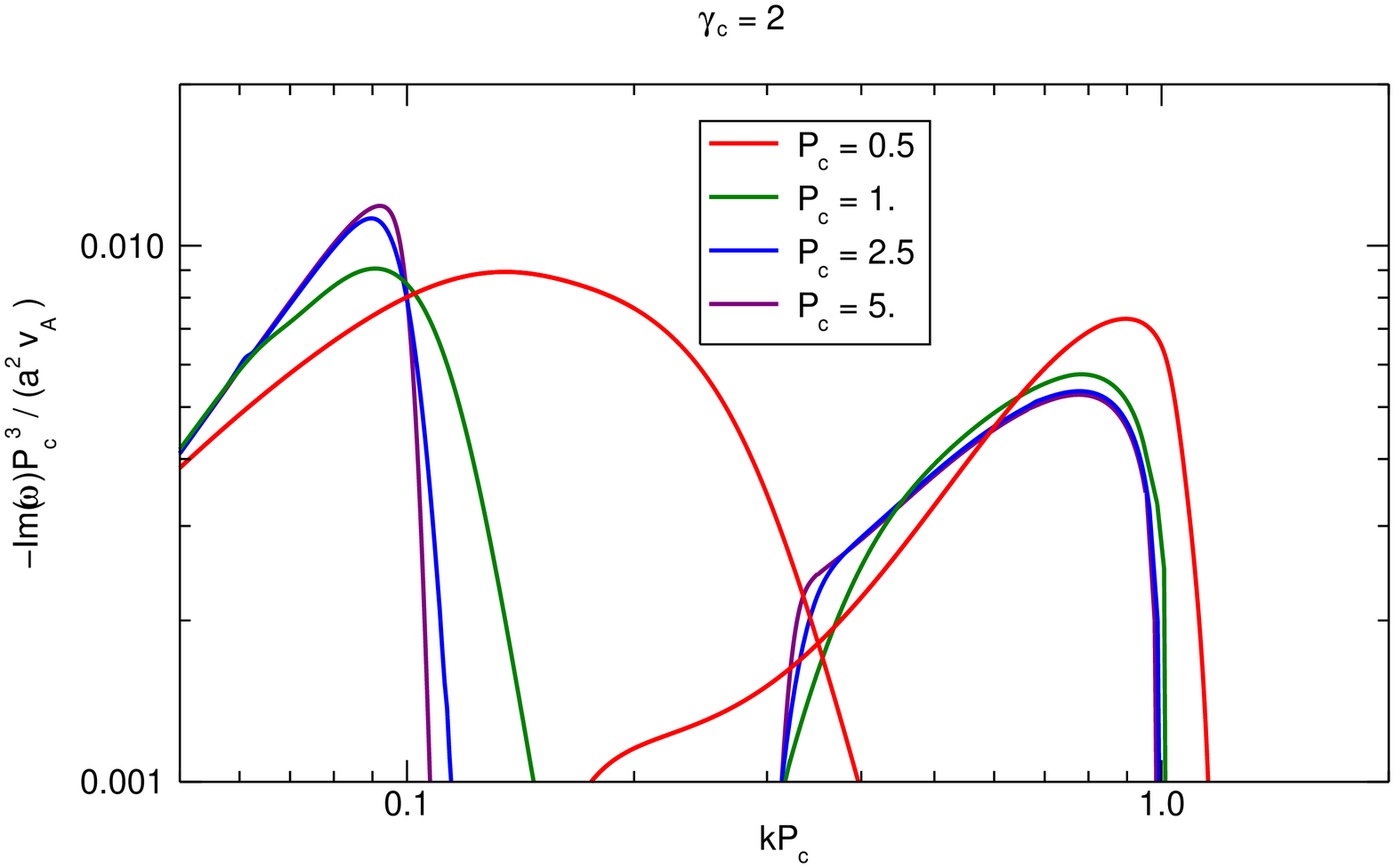} 
  \includegraphics[width=9cm]{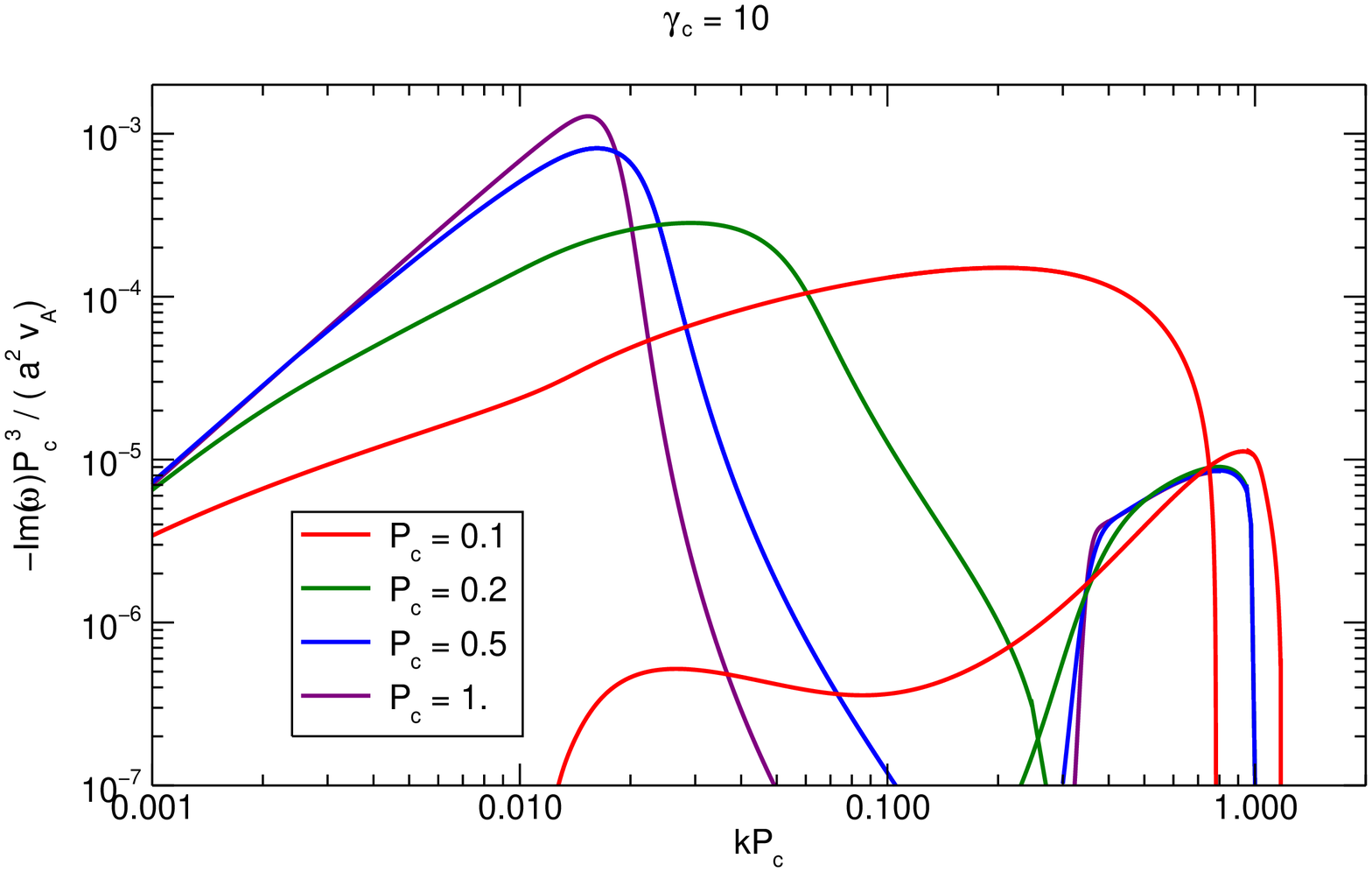} 
  \ \caption{\small Plots of the (normalized) growth rate as a function of the wavenumber, for the four different values of the pitch parameter $P_c$ given in the legends and for $M_a = 0.01$. 
   The three panels refer to three different values of $\gamma_c$, more precisely $\gamma_c = 1.01$ for the top panel, $\gamma_c  = 2$ for the middle panel and $\gamma_ z = 10$ for the bottom panel. 
   The values of $P_c$ are chosen so that in the rest frame we have always $P'_c = 1, 2, 5, 10$. 
   In the relativistic case we observe the splitting of the mode in two branches as described in the text. }
   \label{fig:cd}
\end{figure}
The CD mode is present only for $m=1$ (as in the Newtonian case) and, by lowering $P_c$, it progressively shifts towards larger values of the wavenumber and increases also its growth rate. 
Moreover, for $P_c = 1$, we observe a splitting of the CDI in two unstable regions (see the lower half in the top left panel of Fig. \ref{fig:P10RMHD}). 
This mode splitting can be better understood by inspecting Fig. \ref{fig:cd} where we plot the normalized growth rate as a function of the wavenumber $k$ for $M_a=0.01$ and different values of $P_c$. 
Since for $M_a \lesssim 0.1$ the CDI becomes essentially independent of $M_a$, the curves plotted in Fig. \ref{fig:cd} are representative of the instability behavior for small values of $M_a$.  
The three panels refer to different values of $\gamma_c$, namely, $\gamma_c =1.01$ in the upper panel, $\gamma_c = 2$ in the middle panel and $\gamma_c=10$ in the lower panel. 
In the upper panel we see that the behavior of CDI at small non-relativistic velocities is essentially the same as in the static case (see Fig. \ref{fig:staticmhd} for comparison).

\begin{figure}
   \centering
   \includegraphics[width=9cm]{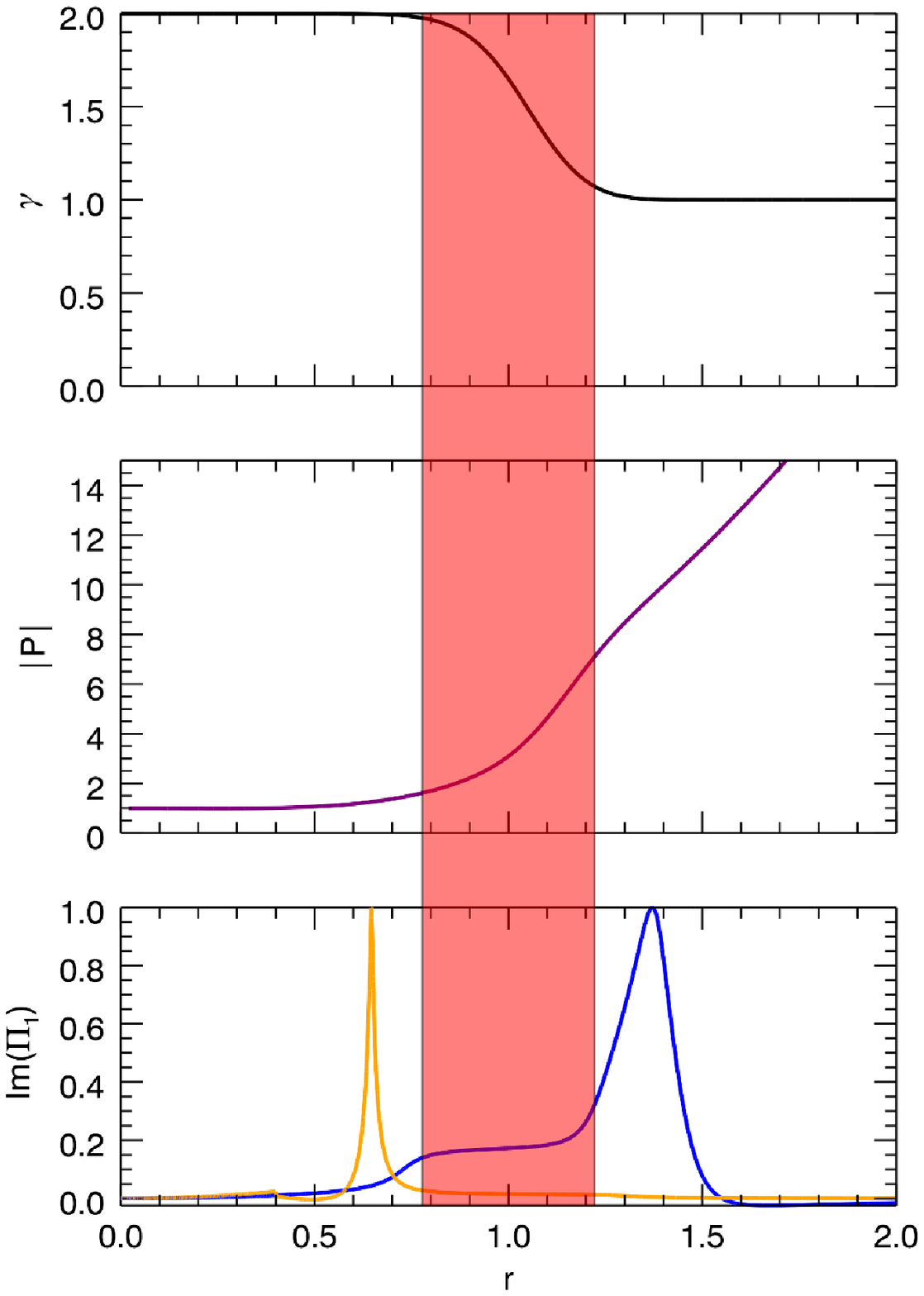} 
   \caption{\small In the top panel we plot the profile of the Lorentz factor $\gamma(r)$,  in the middle panel we plot the pitch profile $P(r)$ and in the bottom panel we plot  the eigenfunctions for the inner (orange) and outer (blue) modes, for $\gamma_c = 2$, $P_c = 1$. 
   The orange curve is for $k=0.8$ (inner mode) and the blue curve is for $k=0.1$ (outer mode).}
   \label{fig:eigen}
\end{figure}
When $\gamma_c$ is increased to $2$ (middle panel), we observe that the mode splits into two.  
In order to comprehend the reason behind the observed mode splitting, we plot in Fig. \ref{fig:eigen} the radial profiles of the Lorentz factor (top panel), pitch (middle panel) and the eigenfunctions relative to the electromagnetic pressure perturbation for two different values of the wavenumber, in the case with $\gamma_c = 2$ and $P_c = 1$ (bottom panel).
The red shading marks the shear region where the jet velocity decreases and pitch profile becomes the steepest.
In plotting the eigenfunctions, we have chosen the wavenumbers $k = 0.1$ (blue curve) and $k = 0.8$  (orange curve) which correspond to the maximum growth rate of each of the two branches shown by the green curves in the middle panel of Fig. \ref{fig:cd}.  
The modes present a resonant behavior and the peak positions are located at the radii where the condition $k P(r) = 1$ is fulfilled.  
Looking at the form of the eigenfunctions in the bottom panel of Fig. \ref{fig:eigen}, we see that the mode with $k=0.8$ peaks inside the jet core in the flat part of the pitch profile. 
At the opposite, the mode with $k =0.1$ reaches a maximum outside the velocity shear region.  
For this reason, we denote the right branch at large wavenumbers as the ``inner mode'' and the left branch at small wavenumbers as the ``outer mode''. 

Given that the absolute value of the pitch is monotonically increasing with radius, we note that the resonant position, expressed by the condition $kP(r)=1$, has to shift to larger radii as $k$ is decreased.
Thus, if we focus on the inner mode, the resonant position will move from $r\sim 0$ (at large $k$) toward the exterior until, for a lower value of $k$, it will fall inside the shear region where the pitch profile is steeper thereby becoming stabilized.
A further decrease in the wavenumber leads the resonance point outside the velocity shear region, where the pitch slope decreases, thus giving rise to the outer branch of the instability.  
%
%
The outer mode, for the considered parameters, has a growth rate that is always somewhat larger than that of the inner mode and it is the one visible in Fig. \ref{fig:P1RMHD} (upper left panel). 

For smaller values of $P_c$, Fig. \ref{fig:cd} shows that the deviation of the growth rate from the $P_c^3$ scaling (Eq. \ref{eq:scaling}) becomes larger and the outer mode stretches while moving towards larger values of $k$.
At the same time, the inner mode widens while moving towards lower values of $k$, until a region of superposition between the two modes is formed.
This behavior can be attributed to the non-monotonic trend of the pitch for small values of $P_c$, as already shown in see Fig. \ref{fig:pitch}.

\begin{figure}
   \centering
   \includegraphics[width=9cm]{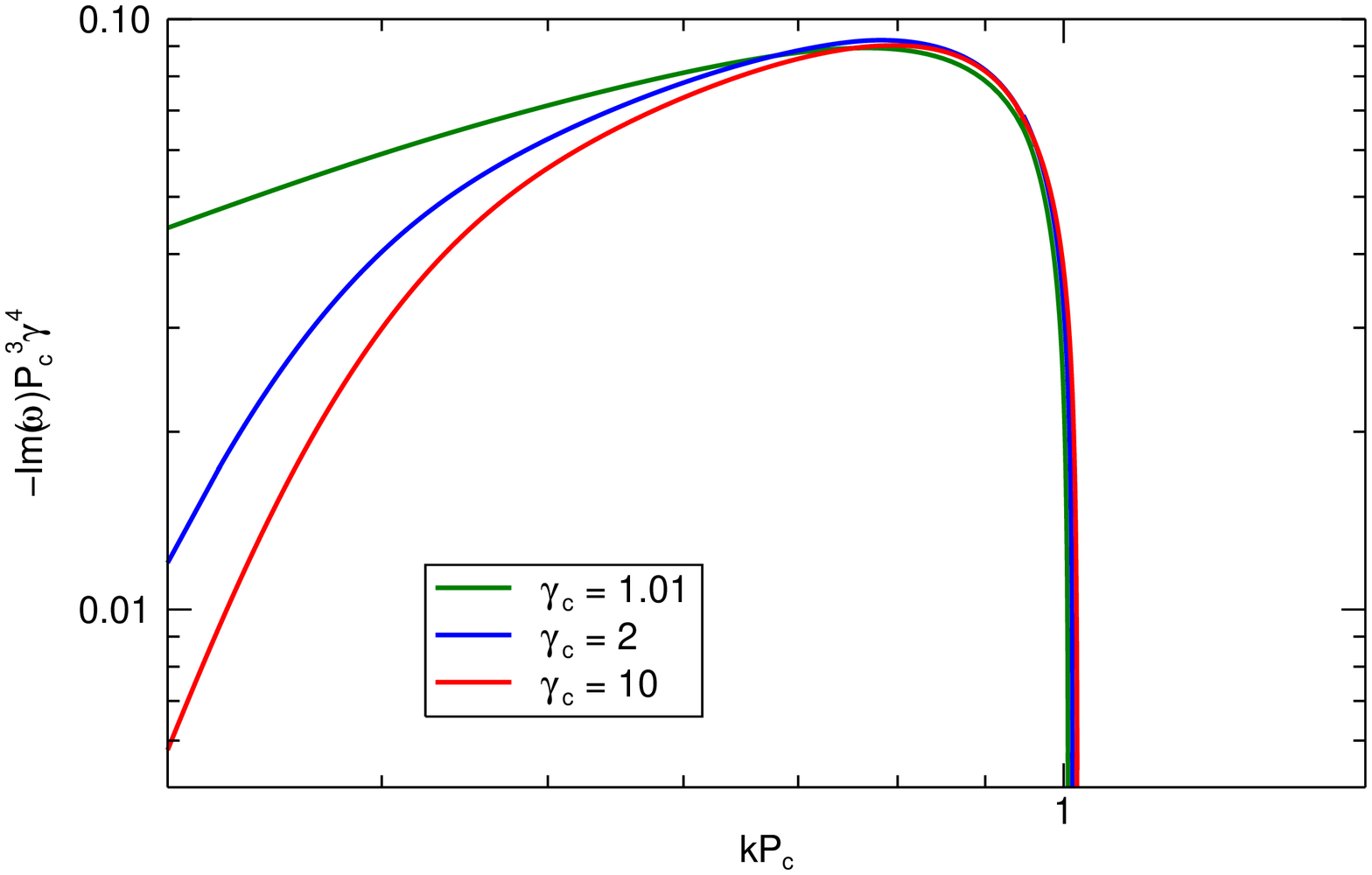} 
   \caption{\small Plot of the growth rate as a function of the wavenumber for three different values of the Lorentz factor $\gamma_c$,  as indicated in the legend. The values of $P_c$ are chosen so that in the jet rest frame the pitch on the axis is equal to 2.  The figure demonstrates the validity of the scaling of the growth rate with $1/\gamma_c^4$ close the maximum. For lower values of the wavenumber the curves deviate from the above scaling.}
   \label{fig:cd_scaling}
\end{figure}

The inner mode, being confined inside the jet core, does not feel the effect of the velocity shear and its properties may be derived by simply applying the appropriate Lorentz transformations to the results obtained in the static case. 
We can in fact relate the pitch, growth rate and wavenumber in the jet frame to those measured in the laboratory frame by the following relations
\begin{equation}\label{eq:Pc_scaling}
  P_c = \frac{P'_c}{\gamma_c}  \,, \qquad  
  \hbox{Im} (\omega) = \frac {\hbox{Im} (\omega')}{\gamma_c} \,, \qquad 
  k = k' \gamma_c  \,,
\end{equation}
where the primed quantities are measured in the jet frame, while the unprimed quantities are measured in the laboratory frame. 
From the scaling given by Eq. ( \ref{eq:scaling}), we obtain
\begin{equation}\label{eq:Im_omega_scaling}
\hbox{Im} (\omega) \sim  \frac{\hbox{Im} (\omega') }{\gamma_c} \; \sim \; \frac{P'^3_c}{\gamma_c}  f(k' P'_c) \; \sim \; \frac{P^3_c}{\gamma_c^4}  f(k P_c) \,.
\end{equation}
The scaling of the growth rate with $1/\gamma_c^4$ is demonstrated by Fig. \ref{fig:cd_scaling}, where we plot $\hbox{Im} (\omega) P_c^3\gamma_c^4$ as a function of the wavenumber for three different values of $\gamma_c$. 
The scaling is excellent around the maximum of the growth rate, where the eigenfunction is more concentrated in the jet core. 
As we move towards smaller values of $k$, the radial extension of the eigenfunction increases, the effect of the velocity shear becomes more important and the three curves deviates from one another. 
Besides, from Eq. (\ref{eq:Pc_scaling}), we have that $k P_c = k' P'_c$ and the stabilization condition in the laboratory frame can be also written as $k P_c = 1$.
   
\begin{figure}
   \centering
   \includegraphics[width=9cm]{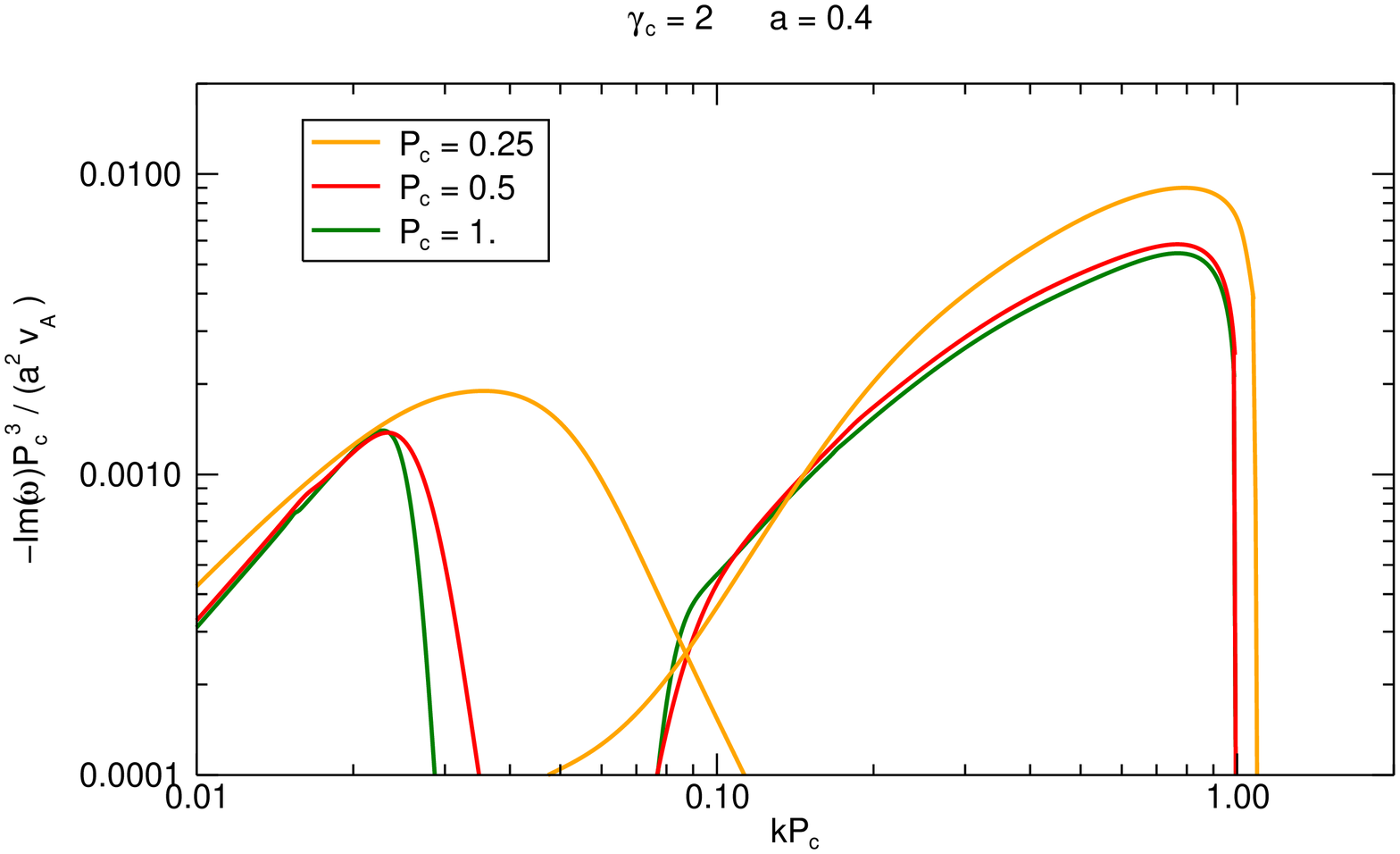} 
   \caption{\small Plot of the growth rate as a function of the wavenumber, for the three different values of the pitch parameter $P_c$ given in the legends and for $M_a = 0.01$ and $\gamma_c = 2$.}
   \label{fig:a03}
\end{figure}
All the results presented so far have been obtained for $a/r_j = 0.6$. 
We recall that the value of the parameter $a$ determines the width of the current distribution and we can then ask how the results depend on it.
The width of the current distribution is related to the extent of the flat part in the pitch profile (see Fig. \ref{fig:pitch}): thus, if we decrease $a$, the pitch grows to larger values inside the jet, the inner mode becomes stabilized at lower wavenumbers and the outer mode is also found at lower wavenumbers. 
The position of the stable region between the inner and outer branches is therefore a function of the value of $a$ and moves towards smaller wavenumbers (when $a$ is decreased) and to larger wavenumbers (when $a$ is increased). 
In addition, the growth rate of the outer branch, being found at small values of $k$, decreases with $a$, while the growth rate of the inner branch is only determined by $P_c$ and does not depend on $a$. 
We then expect that, by decreasing $a$, the inner mode will become dominant. 
As an example, we show in Fig. \ref{fig:a03} how the results obtained for $a = 0.3$ confirm these expectations. 
The figure refers to $\gamma_c = 2$ and can be directly compared to the middle panel of Fig. \ref{fig:cd}.   

We can now summarize the results obtained for the CDI in the relativistic regime: we have two branches of instability, an inner mode concentrated inside the jet and an outer mode in which the perturbation is concentrated outside the jet. 
The relative importance of the two branches depends both on the current concentration determined by the parameter $a$ and on the Lorentz factor $\gamma_c$. 
In fact, as we decrease $a$, the growth rate of the outer branch also decreases and the mode shifts towards smaller wavenumbers, while the growth rate of the inner branch does not change. 
Conversely, as we increase $\gamma_c$, the growth rate of the inner branch decreases while the growth rate of the outer branch does not change.

\section{Summary}
%
%

We have examined the stability properties of a relativistic magnetized cylindrical flow in the approximation of zero thermal pressure, neglecting also the effects of rotation and focusing only on the $|m| = 1$ mode. 
In this configuration we have two kinds of instability that may be present: Kelvin-Helmholtz and current driven. 
The instability behavior depends of course on the chosen equilibrium configuration and this is somewhat arbitrary since we have no direct information on the magnetic field structure, although some indications are provided by the acceleration models \citep[see e.g.][]{Komissarov2007}. 
Nevertheless, the general outcome and the properties of the solutions obtained for a particular configuration, such as the one we adopted here, can be considered valid for a more general class of equilibria. 
Our results can then be considered representative of an equilibrium configuration characterized by a distribution of current concentrated in the jet, with the return current assumed to be mainly found at very large distances.

We can summarize our results by considering the behavior of the system for different values of the ratio between matter and magnetic energy densities.  
For matter dominated flows, the dominant instability is KH and the wavenumber corresponding to the maximum growth rate as well as the growth rate itself scale both as $1/M_a v_c$. Somewhat above equipartion KHI reaches its maximum growth rate and then it becomes rapidly stabilized. Below this stabilization limit  the dominant instability becomes CDI

The dependence of the KHI on the value of the pitch is relatively weak and only for the smallest value of the axial pitch ($P_c$), in the relativistic case, we observe a displacement of the stability limit towards lower values of $M_a v_c$ and a merging with CDI. 
CDI are therefore prevailing for flows in equipartition or magnetically dominated and  the dependence on $M_a v_c$ in these regimes is quite weak. 
The wavenumber corresponding to the maximum growth rate scale as $1/P_c$, the growth rate itself increases with decreasing $P_c$. and the modes have a resonant character with a peak in the eigenfunction at the radial position where $k P(r) = 1$.

At low jet velocity, our equilibrium has  no  return current inside the domain,  while at relativistic velocities we have a small portion of the return current corresponding to the velocity shear region. The corresponding steepening of the pitch profiles induces a stabilization of the modes for which the resonance condition corresponds to radial positions where the return current is found.   
We then observe a splitting of the CDI in two branches, one at high wavenumbers (the inner mode) characterized by an eigenfunction with a resonant peak in the inner radial part of the flow, and one at smaller wavenumbers (the outer mode) for which the resonant peak is outside the jet region.  
Which of the two branches is dominant depends on the width of the current distribution  and on the Lorentz factor of the flow.  
An increase in the current concentration (small $a$) favors the growth of the inner mode, while an increase in $\gamma_c$ enhances the development of the outer mode. 

The different behavior in the explored parameter ranges may have crucial implications for the nonlinear stages as distinct  types of instability may evolve differently. This study is therefore an essential first step for the interpretation of the results of numerical simulations that will be presented in a following paper and for their comparison with astrophysical data.

\section*{Acknowledgments}

\appendix
\section{Derivation of the linearized equations}
\label{ap:linear}
We start from the linearized system (\ref{eq:lincont} - \ref{eq:linMHD}) and,  assuming the perturbations to be of 
the form $\propto \exp \left( {{\mathi}} \omega t - {{\mathi}} m \varphi - {{\mathi}} k z \right)$, rewriting the vectorial equations in 
components, substituting the condition $\nabla \cdot \vec{B_1} = 0$ to the $z$ component of the induction equation,  we obtain the following mixed system of $11$ differential and algebraic equations in the $11$ unknowns $\rho_1, v_{1r}, v_{1\varphi}, v_{1z}, B_{1r}, B_{1\varphi}, B_{1z}, E_{1r}, E_{1\varphi}, E_{1z}, \Pi_1$., where we introduced the total electromagnetic pressure perturbation $\Pi_1$:
\begin{equation}
{\mathi}
\tilde{\omega}\frac{\rho_1}{\rho_0}+\frac{1}{r\rho_0\gamma_0}\frac{d}{dr}(r\rho_0\gamma_0v_{1r})+{\mathi}
\left(\tilde{\omega}\gamma_0^2v_{0\varphi}-\frac{m}{r}\right)v_{1\varphi}+{\mathi}
(\tilde{\omega}\gamma_0^2v_{0z}-k)v_{1z}=0.
\label{eq:app_lin_cont1}
\end{equation}

\begin{eqnarray}
\left. {\mathi}\tilde{\omega}\rho_0\gamma_0^2v_{1r}- \frac{2
\rho_0 \gamma^2_0 v_{0 \varphi}}{r}\left(\gamma^2_0 v_{0
\varphi}^2 + 1 \right) v_{1 \varphi}-\frac{2\rho_0\gamma^4_0
v^2_{0 \varphi} v_{0 z}}{r} v_{1 z} - \frac{\gamma_0^2
v_{0\varphi}^2}{r}\rho_1 \right. = & & \nonumber
\\=-\frac{d\Pi_1}{dr}-\frac{2\Pi_1}{r}+\frac{2B_{0z}B_{1z}}{r}-{\mathi}
\overline{\omega}B_{0z}E_{1\varphi}-{\mathi} k_B B_{1r}+{\mathi} \Omega
r k B_{0z} E_{1z}+{\mathi} \omega B_{0\varphi} E_{1z},
\label{eq:app_lin_momr}
\end{eqnarray}

\begin{multline}
{\mathi} \tilde{\omega} \rho_0 \gamma^2_0
\left(\gamma^2_0v_{0\varphi}^2+1\right)v_{1\varphi}+{\mathi}
\tilde{\omega}\rho_0\gamma_0^4v_{0\varphi}v_{0z}v_{1z}+\frac{\gamma_0\rho_0v_{1r}}{r}\frac{d}{dr}(r\gamma_0v_{0\varphi})=\\
=\frac{B_{1r}}{r}\frac{d}{dr}(rB_{0\varphi})+\frac{{\mathi}
m}{r}B_{0z}B_{1z}-{\mathi} kB_{0z}B_{1\varphi}+{\mathi} \omega
B_{0z}E_{1r}+\frac{E_{1\varphi}}{r}\frac{d}{dr}(rE_{0r})
\label{eq:app_lin_momphi}
\end{multline}

\begin{multline}
{\mathi} \tilde{\omega}\rho_0\gamma^2_0\left(\gamma^2_0
v_{0z}^2+1\right) v_{1z} + {\mathi} \tilde{\omega}
\rho_0\gamma_0^4v_{0\varphi}v_{0z}v_{1\varphi}+\rho_0\gamma_0v_{1r}\frac{d}{dr}(\gamma_0v_{0z})=\\
=B_{1r}\frac{dB_{0z}}{dr}-\frac{{\mathi}
m}{r}B_{0\varphi}B_{1z}+{\mathi} k B_{0\varphi}B_{1\varphi}-{\mathi}
\omega B_{0\varphi}E_{1r}+\frac{E_{1z}}{r}\frac{d}{dr}(rE_{0r}).
\label{eq:app_lin_momz}
\end{multline}

\begin{equation}
\omega B_{1r}=\frac{m}{r}E_{1z}-kE_{1\varphi},
\label{eq:app_lin_indr}
\end{equation}

\begin{equation}
{\mathi} \omega B_{1\varphi}=\frac{dE_{1z}}{dr}+{\mathi} k E_{1r}
\label{eq:app_lin_indphi}
\end{equation}

\begin{equation}
\frac{1}{r}\frac{d}{dr}(rB_{1r})-\frac{{\mathi}
m}{r}B_{1\varphi}-{\mathi} kB_{1z}=0.
\label{eq:app_lin_divB1}
\end{equation}

\begin{equation}
E_{1r}=B_{0\varphi}v_{1z}-B_{0z}v_{1\varphi}+v_{0z}B_{1\varphi}-v_{0\varphi}B_{1z}
\label{eq:app_lin_MHDr}
\end{equation}

\begin{equation}
E_{1\varphi}=v_{1r}B_{0z}-v_{0z}B_{1r}
\label{eq:app_lin_MHDphi}
\end{equation}

\begin{equation}
E_{1z}=-v_{1r}B_{0\varphi}+v_{0\varphi}B_{1r}.
\label{eq:app_lin_MHDz}
\end{equation}

\begin{equation}
\Pi_1=\vec{B}_0\cdot\vec{B}_1-\vec{E}_0\cdot\vec{E}_1=B_{0\varphi}B_{1\varphi}+B_{0z}B_{1z}-E_{0r}E_{1r}
\label{eq:app_lin_P1}
\end{equation}

where we defined

\begin{equation}
\overline{\omega}\equiv\omega-m\Omega,~~
k_B\equiv\frac{m}{r}B_{0\varphi}+kB_{0z},~~
\tilde{\omega}\equiv\omega-\frac{m}{r}v_{0\varphi}-kv_{0z}=\overline{\omega}-\kappa k_B,
\end{equation}

Furthermore, following
\cite{Pariev96}, it is convenient to introduce the radial
displacement of fluid elements, $\xi_{1r}=-{\mathi} v_{1r}/\tilde{\omega}$ and use it instead of $v_{1r}$.  We will now try to express all the variable in terms of $\xi_{1r}$ and $\Pi_1$ and substitute them in Eqs. (\ref{eq:app_lin_momr}) and (\ref{eq:app_lin_divB1} obtaining a system of two first order differential equations in the two unkowns  $\xi_{1r}$ and $\Pi_1$.

From equations (\ref{eq:app_lin_indr}), (\ref{eq:app_lin_MHDphi}) and (\ref{eq:app_lin_MHDz}), we solve for $B_{1r}$,
$E_{1\varphi}$ and $E_{1z}$ in terms of $\xi_{1r}$
\begin{equation}
B_{1r}=-{\mathi} k_B \xi_{1r}
\end{equation}
\begin{equation}
    E_{1 \varphi}  =  {\mathi} \bar{\omega} B_{0z}  \xi_{1r}
\end{equation}
\begin{equation}
    E_{1 z} =  - {\mathi} \left(\Omega k_B r + \bar{\omega} B_{0 \varphi} \right) \xi_{1r},
    \label{eq:app_E1z}
\end{equation}

For further use it is convenient to define the velocity components parallel and transversal
to the background magnetic field
\begin{equation}
u_{1\parallel} = B_{0\varphi}v_{1\varphi} + B_{0z}v_{1z},
~~~~v_{1\varphi}=\frac{B_{0\varphi}}{B_0^2}u_{1\parallel}-\frac{B_{0z}}{B_0^2}u_{1\bot}
\label{eq:app_u1par}
\end{equation}
\begin{equation}
u_{1\bot} =B_{0\varphi}v_{1z} -B_{0z} v_{1\varphi} ,~~~~v_{1z}=\frac{B_{0z}}{B_0^2}u_{1\parallel}+\frac{B_{0\varphi}}{B_0^2}u_{1\bot},
\label{eq:app_u1perp}
\end{equation}
where $B_0^2=B_{0\varphi}^2+B_{0z}^2$.  

Substituting $E_{1z}$ from equation (\ref{eq:app_E1z}) and $E_{1r}$ from
(\ref{eq:app_lin_MHDr}) into equation (\ref{eq:app_lin_indphi}) and eliminating the radial derivative
$dB_{1r}/dr$ from equation (\ref{eq:app_lin_divB1}) , we get
\begin{equation}
k_Bu_{1\bot}-\tilde{\omega}B_{0z}B_{1\varphi}+\tilde{\omega}B_{0\varphi}B_{1z}=FB_{0z}\xi_{1r},
\label{eq:app_alg1}
\end{equation}
where
\[
F\equiv
rk_B\frac{d\Omega}{dr}+\overline{\omega}\left(\frac{dB_{0\varphi}}{dr}-\frac{B_{0\varphi}}{r}-\frac{B_{0\varphi}}{B_{0z}}\frac{dB_{0z}}{dr}\right).
\]
A second relation  comes from the definition of the
electromagnetic pressure, $\Pi_1$,  Eq. (\ref{eq:app_lin_P1}), if we substitute into
it $E_{0r}=-\Omega rB_{0z}$ and $E_{1r}$ from equation
(\ref{eq:app_lin_MHDr}):
\begin{equation}
\Omega r B_{0z}u_{1\bot}+(B_{0\varphi}+\Omega r
B_{0z}v_{0z})B_{1\varphi}+B_{0z}(1-\Omega rv_{0\varphi})B_{1z}=\Pi_1.
\label{eq:app_alg2}
\end{equation}

Now expressing $v_{1\varphi}$ and $v_{1z}$ through $u_{1\parallel}$
and $u_{1\bot}$ from equations (\ref{eq:app_u1par}) and (\ref{eq:app_u1perp}), substituting into
equations (\ref{eq:app_lin_momphi}) and (\ref{eq:app_lin_momz}) and then eliminating $u_{1\parallel}$, we get
the third relation:
\begin{equation}
\tilde{\omega}Yu_{1\bot}+\sigma (k-\omega
v_{0z})B_{1\varphi}+\sigma \left(\omega
v_{0\varphi}-\frac{m}{r}\right)B_{1z}=G\xi_{1r},
\label{eq:app_alg3}
\end{equation}
where
\[
\sigma =\frac{B_0^2}{\rho_0\gamma_0^2},~~~~~
Y=-\frac{\omega}{\tilde{\omega}}\sigma -\frac{B_0^2}{B_0^2-E_0^2},~~~~~
G=W-\frac{\tilde{\omega}E_0(\vec{v}_0\cdot\vec{B}_0)}{B_0^2-E_0^2}H
\]
\[
H=\frac{1}{\gamma_0}\left(B_{0\varphi}\frac{d}{dr}(\gamma_0v_{0\varphi})+B_{0z}\frac{d}{dr}(\gamma_0v_{0z})+\frac{\gamma_0B_{0\varphi}v_{0\varphi}}{r}
 \right)+\frac{k_Bv_{0\varphi}^2}{r\tilde{\omega}}
\]
\begin{multline*}
W=\frac{\tilde{\omega}}{\gamma_0}\left(B_{0\varphi}\frac{d}{dr}(\gamma_0v_{0z})-B_{0z}\frac{d}{dr}(\gamma_0v_{0\varphi})-\frac{\gamma_0B_{0z}v_{0\varphi}}{r}\right)+\nonumber \\
+\sigma \left[(\nabla \cdot {\vec
E_0})\left(\bar{\omega}+\frac{\Omega r B_{0\varphi} k_B}{B_0^2}
\right) - \frac{\vec
{J}_0\cdot\vec{B}_0}{B_0^2}k_B\right].
\end{multline*}
Equations (\ref{eq:app_alg1}), (\ref{eq:app_alg2}) and (\ref{eq:app_alg3}) form a system of three linear
equations
\[
k_Bu_{1\bot}-\tilde{\omega}B_{0z}B_{1\varphi}+\tilde{\omega}B_{0\varphi}B_{1z}=FB_{0z}\xi_{1r},
\]
\begin{equation}
\Omega r B_{0z}u_{1\bot}+(B_{0\varphi}+\Omega r
B_{0z}v_{0z})B_{1\varphi}+B_{0z}(1-\Omega rv_{0\varphi})B_{1z}=\Pi_1
\end{equation}
\[
\tilde{\omega}Yu_{1\bot}+\sigma (k-\omega
v_{0z})B_{1\varphi}+\sigma \left(\omega
v_{0\varphi}-\frac{m}{r}\right)B_{1z}=G\xi_{1r},
\]
from which one can solve for $u_{\bot}, B_{1\varphi}$ and $B_{1z}$
in terms of $\xi_{1r}$ and $\Pi_1$. The solubility of this system
depends on the determinant of the left hand side
\begin{equation}
D=\left|\begin{matrix}
k_B & -\tilde{\omega}B_{0z} & \tilde{\omega}B_{0\varphi}  \\ \\
\Omega r B_{0z} & (B_{0\varphi}+\Omega r B_{0z}v_{0z}) &
B_{0z}(1-\Omega r v_{0\varphi})\\ \\
\tilde{\omega}Y &  \sigma (k-\omega v_{0z}) & \sigma \left(\omega
v_{0\varphi}-\frac{m}{r}\right)\end{matrix}\right|,
\end{equation}
which after simplification reduces to
\begin{equation}
D=(\sigma +1)B_0^2\tilde{\omega}^2+\sigma k_B\left[2\tilde{\omega}(\vec
{v}_0\cdot\vec{B}_0)-\frac{k_B}{\gamma_0^2}\right].
\label{eq:determinant}
\end{equation}

The possible singularities deriving from this determinant are discussed in the main text in Section \ref{sec:linequations}. 
At this point we are able to express all the variables in terms of  
$\xi_{1r}, \Pi_1$ (in particular for expressing $\rho_1$ we make use of Eq. (\ref{eq:app_lin_cont1})), and substituting them into
equations (\ref{eq:app_lin_momr}) and (\ref{eq:app_lin_divB1})  after a long but straightforward algebra, we
arrive at the system of two first order differential equations in
the radial coordinate for the two basic variables -- the radial
displacement and the perturbed electromagnetic pressure:
\begin{equation}
\left. D\frac{d\xi_{1r}}{dr}=\left(C_1+\frac{C_2-D
k_B'}{k_B}-\frac{D}{r}\right)\xi_{1r} - C_3 \Pi_1 \right.
\end{equation}

\begin{multline}
D\frac{d\Pi_1}{dr} = \left[A_1
D-\frac{\rho_0\gamma_0^2v_{0\varphi}^2}{r}\left(C_1+\frac{C_2-D
k'_B}{k_B}\right)\right]\xi_{1r}+
\\+
\frac{1}{r}\left(\rho_0\gamma_0^2v_{0\varphi}^2C_3-2D\right)\Pi_1+\frac{2B_{0z}}{r}D
B_{1z}+A_2D u_{1\bot}
\end{multline}
where $k'_B\equiv dk_B/dr$ and the long expressions for $A_1, A_2,
C_1, C_2, C_3$ and for $Du_{1\bot}, DB_{1z}$ through $\xi_{1r}, \Pi_1$
are given in the Appendix \ref{ap:coefficients}.  

\section{Coefficients of the linear system}
\label{ap:coefficients}
\begin{eqnarray}
A_1\equiv\bar{\omega}^2B_0^2+\Omega^2 r^2 k_B^2 +
2\bar{\omega}B_{0\varphi}\Omega r k_B
-k_B^2+\rho_0\gamma_0^2\tilde{\omega}^2-\frac{\gamma_0v_{0\varphi}^2}{r\tilde{\omega}}\frac{d}{dr}(\gamma_0\rho_0\tilde{\omega})-\nonumber
\\ -\frac{\rho_0v_{0\varphi}H}{r(B_0^2-E_0^2)}\left(2B_{0\varphi}+\gamma_0^2v_{0\varphi}(\vec{v}_0\cdot\vec{B}_0)+\frac{v_{0\varphi}k_B}{\tilde{\omega}}\right),
\end{eqnarray}
\begin{equation}
A_2\equiv\frac{\rho_0\gamma_0^2v_{0\varphi}}{rB_0^2}\left[\frac{v_{0\varphi}}{\tilde{\omega}}\left(kB_{0\varphi}
   -\frac{m}{r}B_{0z}\right)-2B_{0z}+\frac{E_0v_{0\varphi}B_0^2}{B_0^2-E_0^2}-\frac{E_0(\vec{v}_0\cdot\vec{B}_0)}{B_0^2-E_0^2}
     \left(2B_{0\varphi}+\frac{v_{0\varphi}k_B}{\tilde{\omega}}\right)\right],\\
\end{equation}
\begin{align}
C_1&\equiv G\left(\frac{m}{r}B_{0z}-kB_{0\varphi}-\Omega r
B_{0z}\omega\right)-\frac{F\tilde{\omega}B_0^2\Omega r
v_{0z}B_{0z}}{B_0^2-E_0^2},\\
C_2&\equiv FB_{0z}\left[\left(\frac{m}{r}B_{0z}-kB_{0\varphi}
\right)\left(\omega
\sigma +\frac{\tilde{\omega}B_0^2}{B_0^2-E_0^2}\right)-\sigma \Omega r
B_{0z}\left(k^2+\frac{m^2}{r^2}\right)-\frac{mr\tilde{\omega}B_{0z}B_0^2\Omega^2}{B_0^2-E_0^2}\right],\\
C_3&\equiv
\sigma \left(\omega^2-\frac{m^2}{r^2}-k^2\right)+\frac{\tilde{\omega}^2B_0^2}{B_0^2-E_0^2}
\end{align}
and $u_{1\bot}$ and $B_{1z}$ expressed through $\xi_{1r}$ and $\Pi_1$
are
\begin{multline}
D u_{1\bot}=\left[\sigma B_{0z}F\left(\tilde{\omega}(\vec
{v}_0\cdot\vec{B}_0)-\frac{k_B}{\gamma_0^2}\right)-G\tilde{\omega}(B_0^2-E_0^2)\right]\xi_{1r} + \\
+\sigma \tilde{\omega}\left(kB_{0\varphi}-\frac{m}{r}B_{0z}+\omega\Omega
r B_{0z}\right)\Pi_1,
\end{multline}
\begin{multline}
DB_{1z}=\left[G(k_BB_{0\varphi}-\bar{\omega}E_0B_{0z})+ \right. \\
\left. + FB_{0z}\left(\omega \sigma  B_{0\varphi}+\sigma \Omega r B_{0z}k +\frac{\tilde{\omega}B_0^2(B_{0\varphi}+\Omega r B_{0z}v_{0z})}{B_0^2-E_0^2}\right)\right]\xi_{1r}+ \\
  +   \left[\sigma (\omega\bar{\omega}B_{0z}-kk_B)+\frac{\tilde{\omega}^2B_0^2B_{0z}}{B_0^2-E_0^2}\right]\Pi_1,
\end{multline}
where $D$ is the determinant (\ref{eq:determinant}).

The second equation A25, after expressing $DB_{1z}$ and $Du_{1\bot}$ through $\xi_{1r}$ and $\Pi_1$ takes the form
\begin{multline}
D\frac{d\Pi_1}{dr} = \left[A_1
D-\frac{\rho_0\gamma_0^2v_{0\varphi}^2}{r}\left(C_1+\frac{C_2-D
k'_B}{k_B}\right)  + \frac{C_4}{r} +C_5 \right]\xi_{1r}+
\\+
\left[\frac{1}{r}\left(\rho_0\gamma_0^2v_{0\varphi}^2C_3-2D+C_6\right)+C_7\right]\Pi_1
\end{multline},

where the coefficients $C_4,C_5,C_6,C_7$ are given by
\[
C_4=2B_{0z}G(k_BB_{0\varphi}-\bar{\omega}E_0B_{0z}) + 2FB_{0z}^2\left(\omega \sigma  B_{0\varphi}+\sigma \Omega r B_{0z}k +\frac{\tilde{\omega}B_0^2(B_{0\varphi}+\Omega r B_{0z}v_{0z})}{B_0^2-E_0^2}\right),
\]
\[
C_5=A_2\left[\sigma B_{0z}F\left(\tilde{\omega}(\vec
{v}_0\cdot\vec{B}_0)-\frac{k_B}{\gamma_0^2}\right)-G\tilde{\omega}(B_0^2-E_0^2)\right]
\]
\[
C_6=2B_{0z}\sigma (\omega\bar{\omega}B_{0z}-kk_B)+\frac{2\tilde{\omega}^2B_0^2B_{0z}^2}{B_0^2-E_0^2}
\]
\[
C_7=A_2\sigma \tilde{\omega}\left(kB_{0\varphi}-\frac{m}{r}B_{0z}+\omega\Omega
r B_{0z}\right)
\]

\section{Asymptotic solution at small radii}
\label{ap:small_r}

To find solution of equations (\ref{eq:dxi/dr}) and (\ref{eq:dp/dr}) at small radii, we
calculate the coefficients entering these equations at $r\rightarrow
0$ taking into account that in this limit the equilibrium quantities
$v_{0\varphi}, B_{0\varphi}\propto r$, while $v_{0z}$ and $B_{0z}$
tend to constant values. Thus we have (primes everywhere denote
radial derivative)
\[
\lim_{r\rightarrow 0}k_B=kB_{0z}+mB'_{0\varphi},
~~\lim_{r\rightarrow 0}k'_B=kB'_{0z},
\]
\[
\lim_{r\rightarrow
0}D=B_{0z}^2\tilde{\omega}^2+\sigma (\overline{\omega}^2B_{0z}^2-k_B^2)
\]
\[
\lim_{r\rightarrow
0}A_1=\bar{\omega}^2B_{0z}^2-k_B^2+\rho_0\gamma_0^2\tilde{\omega}^2,
~~\lim_{r\rightarrow
0}A_2=-\frac{\rho_0\gamma_0^2v'_{0\varphi}}{B_{0z}}\left(\frac{mv'_{0\varphi}}{\tilde{\omega}}+2\right)
\]
\[
\lim_{r\rightarrow
0}C_1=-\frac{2mB_{0z}^2}{r}\left(\tilde{\omega}v'_{0\varphi}+\sigma \overline{\omega}\Omega+\frac{\sigma B'_{0\varphi}k_B}{B_{0z}^2}\right)
\]
\[
\lim_{r\rightarrow
0}C_2=mB_{0z}^2\left[k_B\frac{d\Omega}{dr}+\overline{\omega}\left(\frac{B''_{0\varphi}}{2}-\frac{B'_{0\varphi}}{B_{0z}}\frac{dB_{0z}}{dr}\right)\right]
\left(\sigma \overline{\omega}+\tilde{\omega}\right)
\]
\[
\lim_{r\rightarrow 0}C_3=-\sigma \frac{m^2}{r^2}
\]
\[
\lim_{r\rightarrow
0}Du_{1\bot}=2\tilde{\omega}B_{0z}^3\left(\tilde{\omega}v'_{0\varphi}+\sigma \overline{\omega}\Omega
+\frac{\sigma B'_{0\varphi}k_B}{B_{0z}^2}\right)\xi_{1r}-\tilde{\omega}m\sigma B_{0z}\frac{\Pi_1}{r}
\]
\begin{multline*}
\lim_{r\rightarrow
0}DB_{1z}=-2B_{0z}(k_BB'_{0\varphi}+\overline{\omega}\Omega B_{0z}^2)
\left(\tilde{\omega}v'_{0\varphi}+\sigma \Omega
\overline{\omega}+\frac{\sigma B'_{0\varphi}k_B}{B_0^2}
 \right)\xi_{1r}+\\+[\sigma (\omega\overline{\omega}B_{0z}-kk_B)+\tilde{\omega}^2B_{0z}]\Pi_1.
\end{multline*}
Substituting these coefficients into equations (\ref{eq:dxi/dr}) and (\ref{eq:dp/dr}), to leading order, we
obtain
\begin{equation}
\frac{d\xi_{1r}}{dr}=-\frac{1}{r}\left[1+\frac{2mB_{0z}^2}{D}\left(\tilde{\omega}v'_{0\varphi}+
\sigma \overline{\omega}\Omega+\frac{\sigma B'_{0\varphi}k_B}{B_{0z}^2}\right)\right]\xi_{1r}+\frac{m^2\sigma }{r^2D}\Pi_1,
\label{eq:ap_small_r1}
\end{equation}
\begin{multline}
\frac{d\Pi_1}{dr}=\frac{2mB_{0z}^2}{rD}\left(\tilde{\omega}v'_{0\varphi}+\sigma \overline{\omega}\Omega+\frac{\sigma B'_{0\varphi}k_B}{B_{0z}^2}\right)\Pi_1+\\
+\frac{D}{\sigma }\left[1-\frac{4B_{0z}^4}{D^2}\left(\tilde{\omega}v'_{0\varphi}+\sigma \overline{\omega}\Omega+\frac{\sigma B'_{0\varphi}k_B}{B_{0z}^2}\right)^2\right]\xi_{1r}.
\label{eq:ap_small_r2}
\end{multline}

We look for solutions in the form $\xi_{1r}\propto r^{\alpha},
\Pi_1\propto r^{\alpha+1}$. After substitution of this form into
equations (\ref{eq:ap_small_r1}) and (\ref{eq:ap_small_r2}) we get
\[
\alpha=\pm |m|-1,
\]
but because a solution must be regular at $r=0$ we take only
$\alpha=|m|-1, (|m|\geq 1)$, and after that the ratio
\begin{equation}
\frac{\Pi_1}{\xi_{1r}}=\frac{r}{m\sigma }\left[{\rm
sign}(m)D+2B_{0z}^2\left(\tilde{\omega}v'_{0\varphi}+\sigma \overline{\omega}\Omega+\frac{\sigma B'_{0\varphi}k_B}{B_{0z}^2}\right)\right]_{|r=0}
\end{equation}
This equation together with the choice $\alpha=|m|-1$ serves as our
boundary condition at small radii.

\section{Asymptotic solution at large radii}
\label{ap:large_r}

To find the asymptotic limit of Eqs. (\ref{eq:dxi/dr}) and (\ref{eq:dp/dr}) and their
corresponding solutions at large radii, we notice that the
equilibrium azimuthal velocity, $v_{0 \varphi}$, the vertical
velocity $v_{0z}$ and $\Omega$ decay very quickly (exponentially)
with radius according to Eqs. (\ref{eq:Omega}), (\ref{eq:vz_prof}), (\ref{eq:H2_prof}) and (\ref{eq:vphi_prof})), so we can
put them effectively zero, $v_{0 \varphi}\simeq 0, v_{0z}\simeq 0,
\Omega \simeq 0$ at large radii and hence $\gamma_0=1$. At large
$r$, the equilibrium density and vertical magnetic field are
constant,  while the azimuthal
magnetic field falls off as $B_{0 \varphi} \propto 1/r$, as
follows from Eqs. (\ref{eq:rho_prof}), (\ref{eq:H2_prof}), (\ref{eq:Bz_prof}) and (\ref{eq:Bphi}). 
Taking this into account, the
asymptotic form of each coefficient entering Eqs.  (\ref{eq:dxi/dr}) and (\ref{eq:dp/dr})  was
calculated at large $r\rightarrow \infty$. Then, neglecting
everywhere terms of the order of $O(r^{-3})$ and higher, after a
rather lengthy algebra, we arrive at the following second order
differential equation for the total electromagnetic pressure
perturbation $\Pi_1$
\begin{equation}
\frac{d^2\Pi_1}{dr^2}+\frac{1}{r}\frac{d\Pi_1}{dr}+\left[\frac{\rho_0+B_{0z}^2}{B_{0z}^2}\omega^2-k^2-\left(m^2+\frac{\rho_0
B_{\varphi c}^2\omega^2}{B_{0z}^4} \right)\frac{1}{r^2}
\right]\Pi_1=0,
\end{equation}
which is of the Bessel's equation type. Its solution corresponding to
radially propagating waves that vanish at infinity is the Hankel
function of the first kind $\Pi_1=H^{(1)}_{\nu}(\chi r)$, where
\[
\chi^2=\frac{\rho_0+B_{0z}^2}{B_{0z}^2}\omega^2-k^2,~~~~\nu^2=
m^2+\frac{\rho_0 B_{\varphi c}^2\omega^2}{B_{0z}^4},
\]
with the leading term of the asymptotic expansion at $r\rightarrow
\infty$
\begin{equation}
\Pi_1=H_{\nu}^{(1)}(\chi r)\simeq \sqrt{\frac{2}{\pi \chi
r}}\exp{\left[{\rm i}\left(\chi r-\frac{\nu
\pi}{2}-\frac{\pi}{4}\right)\right ]}.
\label{eq:p1_asympt}
\end{equation}
The complex parameter $\chi$ can have either positive or negative
sign,
\[
\chi=\pm\sqrt{\frac{\rho_0+B_{0z}^2}{B_{0z}^2}\omega^2-k^2}.
\]
Requiring that the perturbations decay at large radii, we choose the
root of $\chi^2$ in Eq. (\ref{eq:p1_asympt}) that has a positive imaginary part, ${\rm
Im}(\chi)>0$. These perturbations are produced within the jet and
hence at large radii should have the character of radially outgoing
waves. This implies that the real parts of $\chi$ and $\omega$
should have opposite signs, ${\rm Re}(\omega){\rm Re}(\chi)<0$
(Sommerfeld condition), in order to give the phase velocity directed
outwards from the jet.

The asymptotic behaviour of the displacement $\xi_{1r}$ can be
readily obtained from $\Pi_1$ again correct to $O(r^{-3})$
\begin{equation}
\xi_{1r}=\frac{\Pi_1}{\omega^2(\rho_0+B_0^2)-k_B^2}\left({\rm
i}\chi-\frac{1}{2r}\right).
\label{eq:xi1_asympt}
\end{equation}
The asymptotic solutions (\ref{eq:p1_asympt}) and (\ref{eq:xi1_asympt}), together with the above
requirements of outgoing waves with decaying amplitides at
$r\rightarrow \infty$, are used as an initial condition at outer jet
boundary in our numerical scheme for finding eigenvalues of
$\omega$, when doing integration backwards, from large to small
radii.

\label{lastpage}

\end{document}